\documentclass[12pt]{article}
\usepackage{lmodern}
\usepackage[english]{babel}
\usepackage{pdflscape, pslatex} % package to enable inclusion of landscape pages
\usepackage{epstopdf}
\usepackage{float}
\usepackage{bm}

\usepackage[square,sort,comma,numbers]{natbib}
\usepackage{times}
\usepackage{adjustbox} % package for adjusting the table sizes
\usepackage{rotating}
\usepackage{setspace}
\usepackage{listings} % Package for inclusion of R-codes
\usepackage{array}
\usepackage{multirow, multicol} % creation of multiple rows
\usepackage{tabulary,ctable,tabularx}
\usepackage{booktabs}
\usepackage{graphicx}
\usepackage{appendix}
\usepackage{enumerate}% Enable creation of tables spanning over a page
\usepackage{times}
\usepackage{subfig}
\usepackage{amsmath,amsfonts,amssymb,amsthm,titling}

\usepackage[a4paper,width=160mm,top=30mm,bottom=30mm]{geometry}
\newcommand{\R}{\mathbb{R}}	
\usepackage{authblk}

\title{Comparison of Confidence Interval Estimators: An Index Approach}

\author{
	Richard Minkah\thanks{R. Minkah thanks the UG-Carnegie NGAA and UG-BANGA Africa projects for providing financial support.}\\	
	Department of Statistics and Actuarial Science, College of Basic and Applied Science
	University of Ghana, Accra, Ghana\\
	\textit{rminkah@ug.edu.gh} \\
	and\\
	Tertius de Wet\\
	Department of Statistics and Actuarial Science, Stellenbosch University,  Stellenbosch,   South Africa\\
}

\date{}

\begin{document}
\def\spacingset#1{\renewcommand{\baselinestretch}%
	{#1}\small\normalsize} \spacingset{1}

\maketitle
%\author{Richard Minkah}

%\noindent\rule[0.5ex]{\linewidth}{0.5pt}
\begin{abstract}
\noindent\rule[0.5ex]{\linewidth}{1pt}
In many statistical problems, several estimators are usually available for interval estimation of a parameter of interest, and hence, the selection of an appropriate estimator is important. The criterion for a good estimator is to have a high coverage probability close to the nominal level and a shorter interval length. However, these two concepts are in opposition to each other: high and low coverages are associated with longer and shorter interval lengths respectively. Some methods, such as bootstrap calibration, modify the nominal level to improve the coverage and thereby allow the selection of intervals based on interval lengths only. Nonetheless, these methods are computationally expensive. In this paper, we propose an index which offers an easy to compute approach of comparing confidence interval estimators based on a compromise between the coverage probability and the confidence interval length. We illustrate that the confidence interval index has range of values within the neighbourhood of the range of the coverage probability, [0,1]. In addition, a good confidence interval estimator has an index value approaching 1; and a bad confidence interval has an index value approaching 0. A simulation study was conducted to assess the finite sample performance of the index. The proposed index is illustrated with a practical example from the literature.\\

\noindent\rule[0.5ex]{\linewidth}{1pt}
\end{abstract}

\newpage
\spacingset{1.45} % DON'T change the spacing!

\section{Introduction}
%\setcounter{equation}{0}

%\vskip .1truein
%\section{Introduction}

Most statistical problems involve the estimation of some unknown parameter, $\theta,$ of a population from an observed sample using an estimator, $\hat{\theta}$ \cite{Banik2010,Gulhar2012,Pires2008,Zaane2012}. In order to provide a complete description of the information in the sample about $\theta,$ a confidence interval is usually constructed. The key concepts associated with confidence intervals are the coverage probability and interval length. The former is the proportion of times the confidence interval encloses $\theta$ under many replications; and the latter refers to the difference between the upper and the lower confidence limits. These two key concepts are related: longer confidence intervals have higher coverage probabilities approaching the nominal level and shorter confidence intervals have lower coverage probabilities. In statistics, one is often faced with a number of confidence intervals for a parameter arising from different estimators or methods of estimation, and a decision has to be made on the ``best" method of estimation. Since these two key concepts are in opposition to each other, that is, better coverage probability goes with weaker length and vice versa, 
it is useful to have some practical way of combining these measures. In this paper, such an easy to compute measure is proposed and applied to two well-known problems as well as a practical example from the literature.

Suppose we have a sample, $\bm{x}=\{x_1,...,x_n\},$ drawn from an unknown distribution function, $F.$ Let $\ell(\alpha;\bm{x})=\left(\ell_L\left(\alpha;\bm{x}\right),\ell_U\left(\alpha;\bm{x}\right)\right)$ be the $\alpha$-level confidence interval for the unknown parameter $\theta$. Also, denote by $L(\alpha),$ the average of the confidence interval length between the upper confidence limit $\ell_U(\alpha;\bm{x})$ and the lower confidence limit, $\ell_L(\alpha;\bm{x}).$ Furthermore, let the coverage probability be given by $\eta(\alpha)=P\left(\theta \in  \ell(\alpha;\bm{x})\right).$ 

The estimation of $\eta(\alpha)$  is an important issue for statisticians and the goal is to obtain a confidence interval estimator with estimated coverage probability (usually referred to as empirical coverage probability), $\hat{\eta}(\alpha),$ equal to the nominal coverage, $1-\alpha$ \cite{Agresti2000,Loh1987,Loh1991}. However, it is often the case that $\hat{\eta}(\alpha)$ is not exactly equal to $1-\alpha.$ A requirement for a good confidence interval estimator is to have a short interval length and a coverage probability equal to or approximately equal to the nominal coverage. As a result, confidence interval estimator selection can be done by a comparison of the intervals' coverage probabilities and lengths. However, this can be subjective especially if several interval estimators are involved and a compromise is sought between coverage probability and interval length.

A handful of methods to overcome the difficulties in comparing confidence interval estimators rely on an adjustment of the interval lengths such that each interval gives a coverage probability close or equal to the nominal level, $1-\alpha.$ In that case, the comparison of the confidence interval estimators can be done using the confidence interval lengths only. Examples of these methods in the literature include bootstrap calibration \cite{Loh1987, Martin1990} and prepivoting \cite{Beran1987, Lee2003}. The basic idea underlying the bootstrap calibration is to obtain $\beta~ (\beta<\alpha)$ such that the resultant interval's coverage probability equals $1-\alpha$ i.e. $\eta(\beta)=1-\alpha$.  Also, prepivoting involves the transformation of the lower (and/or upper)  confidence level(s) by using its estimated bootstrap distribution function. This has an important application in reducing the coverage error of bootstrap confidence intervals. In addition, prepivoting can be iterated and this automatically moves the empirical coverage closer to the desired level, $1-\alpha.$
%%%%%%%%%%%%%%%%%%%%%%%%%%%%%%%%%%%%%%%%%%%%%%%%

However, in practice, these bootstrap-based procedures generally require computationally costly nested bootstraps (i.e. bootstrapping from the bootstrapped data) from the data. For example, in the case of a double bootstrap, the first bootstrap sample needs $B_1$ resamples from the data and then resampling $B_2$ times from each of single bootstrap samples. Thus, the computational cost involves $B_1\times B_2$ samples in addition to the confidence interval calculations. Also, in \cite{Kilian2000}, the authors were constrained in terms of the number of estimators for impulse responses in large Vector Autoregressive Models due to the prohibitive computational cost. Even for the limited confidence interval estimators considered, in cases like the bias-corrected and accelerated bootstrap method, the computing time required for the evaluation of the estimators was over one year. 

Furthermore, \cite{Beran1987} shows that the coverage precision increases with increasing levels of resamples. Thus, the level of resamples can be done until a point where the coverage is approximately equal to $1-\alpha.$ At this stage, the comparison of interval estimators can be done on the interval lengths only. However, in applications, this is limited by the huge computing power and time needed for such levels of bootstrap. As a result, some work has been done to reduce the computational burden involved in the use of these bootstrap based procedures. Among these, \cite{Loh1987} and \cite{Loh1988} proposed a linear and a nonlinear interpolation respectively to reduce the level of bootstrap replications in calibration. Also, \cite{Nankervis2005} provides an algorithm for the double bootstrap, illustrated above, to reduce the $B_1\times B_2$ total resamples to an appreciable level. These algorithms have varying degrees of success in implementation. Nevertheless, in practical application of these methods, a determination is needed of the benefits of higher levels of resamples against the computational cost. 
%%%%%%%%%%%%%%%%%%%%%%%%%%%%%

In this paper, we propose an index which offers a straightforward approach of comparing confidence interval estimators without the use of Monte Carlo simulation or analytical derivations. The index is based on a compromise between the coverage probability and the confidence interval length. The need for such an index arose from a recent very large simulation study of comparing different estimators of tail index in extreme value theory. Running the simulation to obtain a variety of confidence intervals based on the different estimators, was already computationally very intensive. Applying a further computationally intensive calibration or double bootstrap, would have been too costly in terms of computing time and resources. Hence, the proposed index was developed as a computationally inexpensive compromise between coverage probability and confidence interval length.

The rest of the paper is organised as follows. In section \ref{sec1}, the bootstrap calibration method is presented. The proposed confidence interval index is presented in Section \ref{sec2}. In Section 4, we conduct a simulation study to assess the finite sample performance of the proposed index on four popular confidence interval estimators of the mean from a symmetric and skewed distribution. In addition, several confidence interval estimators of the binomial proportion are examined using the index. Section \ref{sec4} deals with an application of the index on a study of the performance of several confidence interval estimators of the coefficient of variation from \cite{Gulhar2012}.  Finally, we present some concluding remarks in Section \ref{sec5}. 

\section{Bootstrap Calibration}\label{sec1}

\cite{Loh1991} catalogs some procedures for generating confidence intervals with improved coverage probabilities. These include Edgeworth expansion (analytical) and bootstrapping (simulation). The author states, with references, that given that the Edgeworth expansion and bootstrap procedures are valid, both produce results that have the same asymptotic error rates. In particular, the bootstrap procedure implements the Edgeworth correction through simulation in an automatic fashion. In view of this, we consider the bootstrap calibration method of \cite{Loh1987} only. 

Let  $x_1,\ldots,x_n$ be a random sample of size $n$ from the distribution function $F.$ We consider the estimation of the $100(1-\alpha)\%$ 2-sided normal-theory confidence interval of the mean, $\theta=\theta(F),$ given by

\begin{equation}
[\theta_L,~\theta_U]=\left[\hat{\theta}+\frac{z_{\alpha/2}\hat{\sigma}}{\sqrt{n}},~\hat{\theta}-\frac{z_{\alpha/2}\hat{\sigma}}{\sqrt{n}} \right].
\end{equation}  
Here, $\hat{\sigma}$ and $z_{\alpha/2}=\Phi^{-1}(\alpha/2)$ are the unbiased estimate of the variance and the quantile of the standard normal distribution, $\Phi,$ respectively.

If $F$ is normally distributed and $n$ large, the estimated coverage probability, $\hat{\eta}(\alpha),$ will be close to $1-\alpha.$ However, for smaller $n$ and non-normal distributions, $\hat{\eta}(\alpha)$ may differ substantially from $1-\alpha.$ The idea of calibration introduced by \cite{Loh1987} is to replace $\alpha$ with $\beta~ (\beta<\alpha)$ such that 
\begin{equation}\label{Cal}
\hat{\eta}(\beta)\approxeq1-\alpha.
\end{equation}
This invariably implies several or possibly infinite search for $\beta,$ satisfying (\ref{Cal}), and each of this searches is accompanied by bootstrapping samples to obtain $\hat{\eta}(\beta).$ Thus, the method seems impractical. %The calibration idea involves nested bootstrap samples in search of an appropriate value of $\beta.$ 
However,  \cite{Loh1987} and \cite{Loh1988} respectively proposed a linear and a smooth nonlinear interpolation, to reduce the level of bootstrap resampling needed for the seemingly infinite search for $\beta$ to be replaced with just one level. Thus, the calibration is obtained by generating $B$ bootstrap replications.

Let ${\bm{x}}^{*}=\{x_{1}^{*},\ldots,x_{n}^{*}\}$ be a bootstrap sample from $\bm{x}=\{x_1,x_2,\ldots,x_n\}.$  Also, let $\hat{\theta}^*=\hat{\theta}(\bm{x}^{*}),~\hat{\sigma}^*=\hat{\sigma}(\bm{x}^{*})$ and  $t^*_j=\sqrt{n}\left(\hat{\theta}^{*}_{j}-\hat{\theta}\right)/\hat{\sigma}^*$ the $t$ statistic computed from the j\textit{th} bootstrap sample. \cite{Loh1991} defines $\hat{\lambda}_j=1-\Phi\left(|t^*_j|\right)$ and $\beta$ is taken as the $\alpha$-quantile of $\left(\hat{\lambda}_1,\ldots,\hat{\lambda}_B\right).$ In addition, the author argues that the calibration method above is equivalent to the bootstrap root method of \cite{Beran1987}.

The implementation of the calibration method leads to intervals with error rates comparable to bootstrap $t,$ and the accelerated bias-corrected percentile method. However, it is known that these confidence interval estimation methods have limitations with respect to coverage probability and interval lengths \cite{Efron1993}.  

\section{The Index}\label{sec2}

We introduce an index which offers a straightforward approach and avoids the computational burden in the bootstrap-based methods in comparing confidence interval estimators. In addition, the index abstracts the information provided by the confidence interval length and coverage probability, thereby making it a standalone value for comparative purposes. The idea behind the proposed index was to obtain a value that is simple, easy to interpret, and takes into account confidence interval length and coverage probability. In addition, the index is expected to have a range within the neighborhood of the desired coverage probability and hence, can easily be reported together (e.g. graphically) for comparative purposes. 

%This sentence can be expanded explaining that it was derived intuitively in order to have an index that is simple, easy to interpret, takes into account confidence interval length and is within the neighbourhood of the desired coverage probability.

Consider $R$ confidence interval estimators and let $\bm{\eta}=\{\eta_1,\ldots,\eta_R\}'$ and $\bm{L}=\{L_1,\ldots,L_R\}'$ denote the vectors of realised coverage probabilities and average interval lengths respectively.

The confidence interval index, $I$, is defined as

\begin{equation} \label{Index1}
I(L_j,\eta_j;\alpha)=k_\alpha\left(1-\frac{1}{2}\left(\frac{1+H(\eta_j;\alpha)}{1+\left(\frac{\eta_j}{1+L_j}\right)}\right)\right),~~  L\geq0,~ 0\leq\eta_j\leq1, ~j=1,\ldots,R,
\end{equation}
where $k_\alpha$ is a constant depending on the significance level, $\alpha.$ Here, $H$ is a loss function which describes the penalty incurred by the deviation of the empirical coverage probability from $1-\alpha$. In this study, we choose $H$ as a simple absolute loss function defined by

\begin{equation}\label{Loss}
H(\eta_j;\alpha)=|1-\alpha-\eta_j|, ~0\le \eta \le 1,~j=1\ldots,R.
\end{equation}
Consequently, using (\ref{Loss}), the scaling parameter is taken as
\begin{equation}\label{k_alpha}
k_\alpha=\frac{4-2\alpha}{3-2\alpha}, 
\end{equation}
to obtain the range of values of $I(L_j,\eta_j;\alpha)$ within the neighbourhood of the desired coverage probability. To derive the range of values of the index, $I(L_j,\eta_j;\alpha),$ we examine the limit at four extreme cases: 

\begin{enumerate}[\hspace{0.5cm}\bfseries I.]
	\item  $L_j\to 0, \eta_j \to 0 ~ \Longrightarrow ~I(L_j,\eta_j;\alpha) \to \frac{k_\alpha\alpha}{2}.$
	\item  $L_j\to\infty, \eta_j \to 0 ~ \Longrightarrow ~I(L_j,\eta_j;\alpha) \to \frac{k_\alpha\alpha}{2}.$
	\item  $L_j\to \infty, \eta_j \to 1-\alpha ~ \Longrightarrow ~I(L_j,\eta_j;\alpha) \to \frac{k_\alpha}{2}.$
	\item  $L_j\to 0, \eta_j \to 1-\alpha ~ \Longrightarrow ~I(L_j,\eta_j;\alpha) \to 1.$
\end{enumerate}

Thus, $I(L_j,\eta_j;\alpha)$ has a range $\left[k_\alpha\alpha/2,~ 1 \right].$  A bad confidence interval estimator (i.e. an interval with low coverage probability and large interval length) corresponds to cases $\textbf{I}$ and $\textbf{II},$ with $I(L_j,\eta_j;\alpha) \to k_\alpha\alpha/2 .$ On the other hand, a good confidence interval estimator  (i.e. case $\textbf{IV}$)  has  $I(L_j,\eta_j;\alpha) \to 1.$ We note that the range of $I(L_j,\eta_j;\alpha)$ can be transformed to the desirable range of the coverage probability, [0,1], via an affine function $f(x)=2x/(2-k_\alpha\alpha)-k_\alpha\alpha/(2-k_\alpha\alpha),$ for increased interpretability. 

From the aforementioned limits, we conclude that generally a higher value of the index means a better confidence estimator of the parameter $\theta.$ That is, such an estimator has coverage probability close to the nominal value and shorter interval lengths. In addition, as the index penalises for deviation from the nominal level and larger interval lengths, estimators with small coverage probabilities and/or large interval lengths generally have smaller confidence interval index. Therefore, in using this index for comparative purposes, the estimator with largest index value will be chosen ahead of the smaller values. In the subsequent sections, we take $\alpha=0.05$ and, thus, $I(L_j,\eta_j)\in [0.034, 1.000],~~j=1,\ldots,R.$ 

We note that, other loss functions can be chosen for this purpose, for example, quadratic, a Huber function \cite{Huber1992}, among others and appropriate values of $k_\alpha$ determined analytically. For example, if we consider the case of the square loss function, $H(\eta_j;\alpha)=\left(1-\alpha-\eta_j\right)^2, ~0\le \eta \le 1,~j=1\ldots,R.$ The value of $k_\alpha$ can be taken as in (\ref{k_alpha}). However, the limits of $I(L_j,\eta_j;\alpha)$ corresponding to cases I, II, III and IV are respectively $k_\alpha\alpha(2-\alpha)/2$,  $k_\alpha\alpha(2-\alpha)/2$, $k_\alpha/2$ and 1. Thus, the range of the resulting $I(L_j,\eta_j;\alpha),$ is $[\alpha(2-\alpha)^2/(3-2\alpha),1].$ In the case of the two loss functions considered, the rationale behind the choice of $k_\alpha,$ is to obtain a range of values of $I(L_j,\eta_j;\alpha)$ in the neighbourhood of the range of the coverage probability for ease of interpretation. Lastly, the effect of the choice of loss function is reflected in the range of values of $I(L_j,\eta_j;\alpha).$

\section{Simulation Study}\label{sec3}

In this section, we study the performance of the confidence interval index, $I,$ through a simulation study. In this regard, we assess the performance of several confidence interval estimators of the mean from a symmetric and skewed (or asymmetric) distributions. In addition, several estimators of the binomial proportion are assessed using $I.$ 

\subsection{Confidence Interval Index for the Mean}\label{CI_Mean}

We present a simulation study on the estimation of the mean from a symmetric and a skewed distribution in the two subsections that follow. In the former, we considered samples generated from a normal distribution and the latter from a lognormal distribution.

To study the behaviour of the estimators, samples of size, 
$n~ (n=10, 50, 100,\\ 200, 500,~1000)$ were generated from a normal or a lognormal distribution with mean, $\mu,$ and variance, $\sigma^2.$ The parameter of interest is the population mean, $\mu$, which is estimated by the sample mean, $\bar{x}$. The 95\% two-sided confidence interval of  $\mu$  was constructed using four different methods, namely, the normal theory interval, the Johnson $t$ interval ( unlike the normal theory interval, adjust for positive and negative skewness in a data set by shifting the endpoints right and left respectively. The Johnson $t$ interval is given by $\left(\bar{x}+ \hat{\kappa}_3/6\sqrt{n}\left(1+2t_{\alpha}^2\right)\right)\pm t_{\alpha}s/\sqrt{n},$ where $\hat{\kappa}_3$ is the estimate of the population skewness $E\left(X-\mu\right)^3/\sigma^3,~~t_\alpha$ is the $\alpha$ quantile of the $t$ distribution with $n-1$ degrees of freedom and $s$ is the sample standard deviation.) \cite{Johnson1978} and the bootstrap-based intervals-the bootstrap percentile and the  Bias-Corrected and accelerated (BCa) \cite[Chapters 12 and 14]{Efron1993}. 

The following procedure was used to compute the index and its summary statistics:

\begin{itemize}
	\item[A1.] Generate $N (N=1000)$ samples each of size $n$  from $N(\mu,\sigma^2).$\label{AL1}
	
	\item[A2.] Draw $B (B=1000)$ bootstrap samples from each sample in A1 and use these to compute $B$ bootstrap confidence intervals (i.e. bootstrap percentile and the BCa) of the mean. Compute the average of the $B$ interval lengths, $L,$ and the empirical coverage probability $\hat{\eta}$ for both bootstrap interval types separately. \label{AL2}
	
	\item[A3.] Compute the confidence interval for the mean using the normal theory interval and Johnson $t$ interval using each of the $N$ samples in A1. Calculate the average of the $N$ interval lengths, $L,$ and the empirical coverage probability $\hat{\eta}$ for the two interval types. \label{AL3} 
	
	\item[A4.] Repeat A1-A3 a large number of times $R (R=5000),$ to obtain the pairs $\{(\hat{\eta}_1,L_1),\ldots,(\hat{\eta}_R,L_R)\}$ and, hence, the confidence interval index, $I^{(i,j)},~i=1,\ldots,4,~j=1,\ldots,R.$\label{AL4}
	
	\item[A5.] Compute summary statistics for the indexes, $I^{(i,.)}~i=1,\ldots,4.$ \label{AL5} 
\end{itemize}

\subsubsection{Mean of a Symmetric Distribution}\label{SecSym}
Table \ref{tab1} shows the summary statistics of the index for the four interval types computed for observations from $N(2,1).$ It can be seen that, as the sample size increases, $I$ tends to 1: the confidence interval estimators improve with increasing sample size. This is expected in line with the weak law of large numbers:  $\bar{x}$ approaches $\mu$ as $n\to \infty$. 
% latex table generated in R 3.2.1 by xtable 1.7-4 package
% Tue Aug 16 01:42:33 2016

\begin{table}[htp!]
	\centering
	\caption{Summary statistics for the Confidence Interval Index of the Mean}
	\begin{tabular}{llcccc}
		\toprule
		&  & \multicolumn{4}{c}{$I$}  \\ \cmidrule{3-6} 
		$n$	&	Basic Statistics & Normal Theory & Johnson $t$ &Bootstrap Percentile  & BCa\\
		\hline
		&Mean & 0.8555 & 0.8600 & 0.8443 & 0.8401  \\ 
		%	&Median & 0.8557 & 0.8606 & 0.8446 & 0.8404  \\ 
		$10$	&Skewness & -0.4151 & -2.0374 & -0.2722 & -0.3029  \\ 
		& Kurtosis & 0.5392 & 8.1841 & 0.1896 & 0.3133 \\ 
		&St. dev & 0.0075 & 0.0033 & 0.0084 & 0.0089 \\ 
		\hline
		&	Mean & 0.8895 & 0.8913 & 0.8816 & 0.8793  \\ 
		%	&	Median & 0.8897 & 0.8921 & 0.8816 & 0.8797 \\ 
		$50$	&	Skewness & -0.4417 & -1.4934 & -0.0191 & -0.2255  \\ 
		&	Kurtosis & -0.3837 & 2.5216 & -0.0874 & -0.0693 \\ 
		&	St. dev & 0.0061 & 0.0031 & 0.0069 & 0.0074 \\ 
		\hline
		&	Mean & 0.9229 & 0.9231 & 0.9179 & 0.9158\\ 
		%	&	Median & 0.9242 & 0.9240 & 0.9185 & 0.9159  \\ 
		$100$	&	Skewness & -1.1576 & -1.5628 & -0.4305 & -0.2758\\ 
		&	Kurtosis & 0.9175 & 2.7156 & -0.1391 & -0.3270\\ 
		&	St. dev & 0.0042 & 0.0029 & 0.0062 & 0.0064\\ 
		\hline
		&Mean & 0.9708 & 0.9709 & 0.9693 & 0.9691 \\ 
		%	&Median & 0.9716 & 0.9716 & 0.9700 & 0.9698\\ 
		$500$&Skewness & -1.6225 & -1.7231 & -0.7722 & -0.8726  \\ 
		&Kurtosis & 2.7522 & 3.2814 & -0.2012 & 0.1894 \\ 
		&St. dev & 0.0021 & 0.0020 & 0.0033 & 0.0033 \\\hline
		& Mean & 0.9777 & 0.9778 & 0.9745 & 0.9732 \\ 
		%	& Median & 0.9786 & 0.9785 & 0.9750 & 0.9740 \\ 
		$1000$&Skewness & -2.1632 & -2.2197 & -1.0131 & -0.7991 \\ 
		& Kurtosis & 6.7939 & 7.2921 & 0.8867 & 0.4052\\ 
		&St. dev & 0.0029 & 0.0029 & 0.0048 & 0.0053 \\ 
		\bottomrule
	\end{tabular}
	\label{tab1}
\end{table}

In addition, for smaller sample sizes (i.e. $n\le50$), the Johnson $t$ interval has the largest $I$ values in most cases followed by the normal interval. Generally, these two estimators provide better confidence intervals as they have larger index values for measures of location, smaller variability, large negative skewness and  large peakedness. In the case of large sample sizes (i.e. $n>50$), there is not much difference between the performance of the normal theory and Johnson $t$ interval estimators of the mean. The bootstrap percentile is the next best confidence interval estimator of the mean followed by the BCa interval estimator based on the summary statistics. Since the sample mean is an unbiased estimator of the population mean, the percentile interval is expected, as shown in the simulation study, to give better intervals in terms of coverage and interval lengths. We remark that the simulation was carried out for larger sample variances and the results show wider interval length leading to smaller values of the index. Due to space consideration, the results are not included but can be obtained from the authors upon request.

%Also, for larger values of $\sigma,$ the corresponding confidence intervals for the mean are wider. This is reflected in the mean of $I$ having a decreasing order in Tables \ref{tab1},  \ref{IntMean1} and \ref{IntMean2} as $\sigma$ increases from 1 through to 3.  Similarly, Tables \ref{IntMean1} and \ref{IntMean2} in Appendix A show the corresponding results for $N(2,4)$ and $N(2,9)$ respectively.

Furthermore, we consider the performance of $I$ in relation to bootstrap calibration of \cite{Loh1987} and \cite{Loh1988}. Again, the estimation of the mean of a normal distribution is considered. Here, we considered smaller sample sizes where the empirical coverage probability tends to be smaller than the nominal level, $1-\alpha.$ In that case, calibration can be used to increase the empirical coverage probability to approximately equal to $1-\alpha.$ Our aim in this case is to assess the conclusions reached for calibrated intervals in relation to the index. The results of the simulation study for observations from $N(2,1)$ are presented in Table \ref{Calibration}. %In addition, the results for observations from $N(2,4)$ and $N(2,9)$ are presented in Tables \ref{Cal1} and \ref{Cal2} respectively in Appendix B.

\begin{table}[htp!]
	\centering
	\caption{Calibrated and non-calibrated interval estimators for the mean}
	\begin{tabular}{lllllllll}
		\toprule
		&	& \multicolumn{3}{c}{Non-calibrated}& &\multicolumn{3}{c}{Calibrated}\\\cmidrule{3-5}\cmidrule{7-9}
		Sample size	& Estimator	& CP & L & $I$  & & CP & L & $I$  \\
		\hline  
		\multirow{3}*{$n=10$}
		&
		Normal theory & 0.907 & 1.196 & 0.849  & & 0.922 & 1.290 & 0.852  \\ 
		&Johnson $t$ & 0.936 & 1.380 & 0.855 & & 0.960 & 1.518 & 0.853 \\ 
		&Bootstrap Percentile & 0.884 & 1.128 & 0.838 & & 0.906 & 1.219 & 0.846  \\ 
		&BCa & 0.889 & 1.146 & 0.840  && 0.906 & 1.235 & 0.845 \\ 
		\hline  
		\multirow{4}*{$n=15$}
		&
		Normal theory  & 0.931 & 0.988 & 0.878 &  & 0.945 & 1.045 & 0.883  \\ 
		&Johnson $t$  & 0.950 & 1.081 & 0.883 & & 0.950 & 1.081 & 0.883 \\ 
		&Bootstrap Percentile & 0.918 & 0.953 & 0.873 & & 0.929 & 1.004 & 0.876  \\ 
		&BCa & 0.917 & 0.959 & 0.872 &  & 0.931 & 1.013 & 0.876  \\ 
		\hline  
		\multirow{4}*{$n=20$}
		&
		Normal theory  & 0.930 & 0.866 & 0.887  && 0.942 & 0.890 & 0.893  \\ 
		&Johnson $t$ & 0.946 & 0.924 & 0.892  && 0.952 & 0.953 & 0.892  \\ 
		&Bootstrap Percentile & 0.922 & 0.842 & 0.884  && 0.931 & 0.865 & 0.888  \\ 
		&BCa & 0.919 & 0.847 & 0.882  && 0.929 & 0.870 & 0.886  \\ 
		\hline  
		\multirow{4}*{$n=30$}
		&
		Normal theory & 0.951 & 0.708 & 0.912 & & 0.951 & 0.708 & 0.912  \\ 
		&Johnson $t$ &  0.962 & 0.739 & 0.907 & & 0.962 & 0.739 & 0.907  \\ 
		&Bootstrap Percentile & 0.950 & 0.695 & 0.914 & & 0.950 & 0.695 & 0.914   \\ 
		&BCa & 0.946 & 0.699 & 0.911 & & 0.946 & 0.699 & 0.911  \\ 
		\hline  
		\multirow{4}*{$n=50$}
		&
		Normal theory  & 0.950 & 0.550 & 0.928  && 0.950 & 0.550 & 0.928  \\ 
		&Johnson $t$ &  0.953 & 0.564 & 0.926 && 0.953 & 0.564 & 0.926\\ 
		&Bootstrap Percentile & 0.941 & 0.545 & 0.923  && 0.958 & 0.575 & 0.923  \\ 
		&BCa & 0.945 & 0.545 & 0.926   && 0.954 & 0.572 & 0.925  \\ 
		\hline
	\end{tabular}
	\label{Calibration}
\end{table}

For smaller sample sizes $(n\le20),$ the Johnson $t$ interval has empirical coverage probabilities close to the nominal level of 0.95. Calibration of such interval leads to overestimation of the coverage probability. Therefore, we failed to calibrate interval estimators with empirical coverage probability close to $0.95.$  The performance of the Johnson $t$ confidence interval estimator is expected as it adjusts for the skewness in the data (in particular for small sample sizes where skewness is prevalent). However, this interval consistently has the largest interval length compared with the normal theory, bootstrap percentile and BCa. The index values for the Johnson $t$ interval are the largest, and thus, can be considered as the most appropriate confidence interval estimator of the mean.

As the sample size increases, the normal theory interval  and the Johnson $t$ interval estimators outperform the other intervals in terms of coverage probability. Also, the normal theory interval estimator has interval lengths fairly competitive to the bootstrap-based intervals and outperforms the Johnson $t$ interval. This can be seen from the index of the normal theory interval having the largest values. 

In general, we note that calibration as demonstrated in Table \ref{Calibration} does not necessarily bring the empirical coverage up to the desired level of $1-\alpha.$ Thus, calibration, although expensive, is not always attractive. We find that the conclusions of the confidence interval index for the non-calibrated intervals agree mostly with that of the calibrated intervals. Again, if the coverage probability is close to the nominal value, calibration leads to overestimation of the coverage probability. However, the index penalises such intervals for the deviation from the nominal level, and thus, discriminates good intervals from the bad ones. 

\subsubsection{Mean of a Skewed Distribution}\label{SecSkew}

%In Section \ref{SecSym}, we considered the confidence interval for the mean of a symmetric distribution.

In this section, we consider the performance of the confidence interval index for the estimation of the mean of a skewed distribution. Observations were generated from a lognormal distribution with mean $0.$ Since the skewness of the lognormal distribution depends only on the variance, we took variances of 3, 2 and 0.2 corresponding to skewness of 23.732, 6.185 and 1.516 respectively. The results for these three values are shown in Tables \ref{skew1}, \ref{skew2} and \ref{skew3} respectively. 

Firstly, for a largely skewed distribution, it is evident that the normal theory interval has the smallest average confidence interval indexes but relatively larger standard deviations. The normal intervals are symmetric, and hence, has a challenge when it is used to provide a confidence interval for the mean of a heavily-skewed distribution. On the other hand, the BCa interval records the best performance as it has the largest average confidence interval indexes. This results from the fact that for a heavily-skewed distribution, large bias is expected but the BCa interval corrects for bias and skewness and hence, provides better intervals that enclose the actual parameter being estimated.

\begin{table}[htp!]
	\centering
	\caption{Summary statistics for the Confidence Interval Index of the Mean from $Lognormal(0,3)$}
	\begin{tabular}{llcccc}
		\toprule
		&  & \multicolumn{4}{c}{$I$}  \\ \cmidrule{3-6} 
		$n$	&	Basic Statistics & Normal Theory & Johnson $t$ &Bootstrap Percentile  & BCa\\
		\hline
		\multirow{3}*{$10$}
		&Mean & 0.4067 & 0.4252 & 0.4091 & 0.4522 \\ 
		%&Median & 0.4067 & 0.4242 & 0.4090 & 0.4508 \\ 
		&Skewness & 0.1708 & 0.1861 & 0.2562 & 0.1138 \\ 
		&Kurtosis & 0.0329 & 0.1244 & 0.0155 & 0.0543 \\ 
		&St. dev & 0.0154 & 0.0152 & 0.0152 & 0.0150 \\ 
		\hline\multirow{3}*{$50$}
		&	Mean & 0.5101 & 0.5142 & 0.5190 & 0.5646 \\ 
		%&	Median & 0.5098 & 0.5140 & 0.5184 & 0.5646 \\ 
		&	Skewness & -0.0834 & -0.0800 & 0.0123 & 0.0611 \\ 
		&	Kurtosis & -0.0276 & 0.0814 & -0.0294 & -0.0544 \\ 
		&	St. dev & 0.0143 & 0.0144 & 0.0146 & 0.0128 \\ 
		\hline\multirow{3}*{$100$}
		&Mean & 0.5485 & 0.5508 & 0.5568 & 0.5952 \\ 
		%	&Median & 0.5493 & 0.5516 & 0.5570 & 0.5951 \\ 
		&Skewness & -0.0954 & -0.1354 & -0.1217 & -0.0438 \\ 
		&Kurtosis & 0.1251 & 0.1313 & 0.3955 & -0.1864 \\ 
		&	St. dev & 0.0134 & 0.0133 & 0.0133 & 0.0122 \\ 
		\hline\multirow{3}*{$500$}
		&Mean & 0.6209 & 0.6215 & 0.6255 & 0.6454 \\ 
		%&Median & 0.6212 & 0.6217 & 0.6256 & 0.6450 \\ 
		&Skewness & 0.0489 & 0.0370 & 0.0436 & 0.1057 \\ 
		&Kurtosis & -0.3218 & -0.3116 & -0.2307 & 0.0324 \\ 
		&St. dev & 0.0112 & 0.0112 & 0.0109 & 0.0105 \\ 
		\hline\multirow{3}*{$1000$}
		
		&Mean & 0.6487 & 0.6491 & 0.6511 & 0.6638 \\ 
		%	&Median & 0.6490 & 0.6492 & 0.6508 & 0.6638 \\ 
		&Skewness & -0.0177 & -0.0281 & 0.1117 & -0.0156 \\ 
		&Kurtosis & -0.1513 & -0.1528 & 0.0376 & -0.2113 \\ 
		&St. dev & 0.0109 & 0.0108 & 0.0113 & 0.0104 \\ 
		\bottomrule
	\end{tabular}
	\label{skew1}
\end{table}

Secondly, in the case of a moderately skewed distribution, the Johnsons-$t$ interval estimator is by far the best estimator of the mean of the lognormal distribution with larger index values. This is followed by the BCa interval estimator especially for smaller sample sizes where skewness is high. However, as the sample size increases, the normal interval estimator improves and surpasses the BCa with larger confidence interval indexes.  

Thirdly,  for the case of low skewness, i.e. $\sigma^2=0.2,$ the performance relatively follows a similar pattern but with some notable differences. The Johnson $t$ remains the best estimator but the performance of the normal theory interval improves significantly for large sample sizes giving large values of the index. At $n=1000,$ there is not much difference between the two estimators. However, the performance of the BCa interval reduces as its index values are smaller compared with the other estimators. This may be attributed to the low skewness of the distribution, and hence, being close to a symmetric distribution similar to the normal distribution presented in Section \ref{SecSym}. 

Lastly, the index values increase with decreasing variance of the lognormal distribution (i.e. decreasing skewness) signifying better performances than for the case of larger variances (i.e. increasing skewness). This is to be expected as the confidence interval estimators give intervals that have better coverage and smaller interval lengths when skewness is small. This is in conformity with earlier studies that compared estimators of the mean of a skewed distribution based on interval length and coverage probability \cite{Banik2010}. Therefore, in general, the confidence interval index works well for selecting confidence interval estimators of the mean of a skewed distribution.

\subsection{The Confidence Interval Index for the Binomial Proportion}

In this section, we consider the estimation of the binomial proportion by sampling from a binomial distribution. This issue usually arises in applied statistics e.g. incidence rates (in medical science), proportion of defective items (in manufacturing), among others. In addition, unlike the confidence interval for the mean, that of the proportion enables us to measure the performance of the index on non-symmetric intervals.  

Assume $X$ is binomially distributed with parameters $n$ and $p,$ written, $X\sim Bin(n,p).$ Here, the estimator, $\hat{p},$ is the maximum likelihood estimator given by $\hat{p}=X/n.$ This estimator is consistent and, since, the expected value of $X$ is equal to $np,$ it is also unbiased. 

The most basic form of an interval estimate for the proportion, $p,$ is the Wald interval, 
\begin{equation}\label{Wald}
\hat{p}\pm Z_{\alpha/2}\sqrt{\hat{p}(1-\hat{p})/n}
\end{equation}
\cite{Leemis1996,Brown2001,Brown2002}. The properties of this interval estimator have been studied extensively in the literature. Its performance is known to be erratic with respect to coverage probability. In addition, recommendations concerning the values of $n$ and $p$ where this interval is appropriate are conflicting \cite{Leemis1996}. Several attempts have been made to obtain better confidence interval estimators of the binomial proportion. For example, \cite{Agresti2000} proposed an amendment of the interval (\ref{Wald}) by defining $\hat{p}$ as $(X+2)/(n+2).$ Some further modifications and other intervals that are not based on the normality assumption are presented in Table \ref{Est_Tab}. 

\begin{table}[htp!]
	\centering
	\caption{Confidence intervals for the binomial proportion}
	\begin{adjustbox}{width=\textwidth}
		\begin{tabular}{lll}
			\toprule
			Estimator     & Confidence Interval       & Reference\\\hline
			
			Exact & $\left(\mbox{Beta}\left(\alpha/2;X,n-X+1\right),~ \mbox{Beta}\left(1-\alpha/2;X,n-X+1\right)\right) $  &\cite{Agresti2000}\\
			Wald & $\hat{p}\pm Z_{\alpha/2}\sqrt{\hat{p}(1-\hat{p})/n}$  & \cite{Brown2001}\\
			Arcsin &$\sin{^2\left(\arcsin\sqrt{\frac{X-1/2}{n}}\pm\frac{Z_{\alpha/2}}{2\sqrt{n}}\right)}$ & \cite{Pires2008}\\
			
			Arcsin.CC &$\sin{^2\left(\arcsin\sqrt{\frac{X-1/8}{n+3/4}}\pm\frac{Z_{\alpha/2}}{2\sqrt{n+1/2}}\right)}$ & \cite{Pires2008}\\
			
			Pois &$\left(\frac{1}{2n}\chi^2_{2X,1-\alpha/2},~ \frac{1}{2n}\chi^2_{2(X+1),1-\alpha/2}\right)$ & \cite{Leemis1996} \\
			
			BCG & $\frac{X+Z_{\alpha/2}^2/2}{n+Z_{\alpha/2}^2}\pm Z_{\alpha/2}\sqrt{\frac{X}{n^2}\left(1-\frac{X}{n}\right)}$  & \cite{Brown2002}\\

			Wils & $\frac{X+Z_{\alpha/2}^2/2}{n+Z_{\alpha/2}^2}\pm \sqrt{\frac{nZ_{\alpha/2}^2}{\left(n+Z_{\alpha/2}^2\right)^2}\left(\frac{X}{n}\left(1-\frac{X}{n}\right)+\frac{Z_{\alpha/2}^2}{4n}\right)}$  & \cite{Wilson1927}\\
			
			Wils.CC & $\left(\frac{2X+Z_{\alpha/2}^2-1-Z_{\alpha/2}\sqrt{Z_{\alpha/2}^2-2-1/n+4X\left(1-X/n+1/n\right)}}{2\left(n+Z_{\alpha/2}^2\right)},~\frac{2X+Z_{\alpha/2}^2+1+Z_{\alpha/2}\sqrt{Z_{\alpha/2}^2+2-1/n+4X\left(1-X/n-1/n\right)}}{2\left(n+Z_{\alpha/2}^2\right)}\right)$ & \cite{Wilson1927}\\

			AgreC & $\frac{X+Z_{\alpha/2}^2/2}{n+Z_{\alpha/2}^2}\pm Z_{\alpha/2}\sqrt{\frac{X+Z_{\alpha/2}^2/2}{\left(n+Z_{\alpha/2}^2\right)^2}\left(1-\frac{X+Z_{\alpha/2}^2/2}{n+Z_{\alpha/2}^2}\right)}$  & \cite{Agresti1998}\\	
			
			Ag.add4 & $\frac{X+2}{n+4}\pm Z_{\alpha/2}\sqrt{\frac{X+2}{(n+4)^2}\left(1-\frac{X+2}{n+4}\right)}$  & \cite{Agresti2000}\\	
			
			mid-P & $\left(\mbox{Beta}\left(\alpha/2;X+1/2,n-X+1/2\right), \mbox{Beta}\left(1-\alpha/2;X+1/2,n-X+1/2\right)\right) $  &\cite{Agresti2000}\\
			\bottomrule
		\end{tabular}
	\end{adjustbox}
	\label{Est_Tab}
\end{table}

In the present study, we generated samples of size, $n,$ and proportion, $p,$ from a binomial distribution. Each estimator in Table \ref{Est_Tab} is used to obtain a confidence interval for $p.$ We repeat the process $R~ (R=1000)$ times and obtain the average confidence interval length and coverage probability. We then compute diagnostic checks on these intervals using the index $I$ in (\ref{Index1}). 

Tables \ref{Prop_Tab1}-\ref{Prop_Tab3} show the results of the simulation for combinations of $n$ and $p.$ 
%In the case of smaller sample size $(n=20),$ the average interval lengths does not exhibit large variation: the range is 0.0476. In addition, the smallest and largest empirical coverage probabilities are 0.888 and 0.996 respectively. As a result, there is not much difference between the values of $I.$ Similar conclusions can be drawn for the various combinations of $n$ and $p.$ 
%\clearpage
%\newpage
% latex table generated in R 3.1.3 by xtable 1.7-4 package
% Mon Aug 15 15:08:06 2016

\begin{table}[htp!]
	\centering
	\caption{Confidence interval index for the binomial proportion using  $n=10$}
	\begin{adjustbox}{width=\textwidth}		
		\subfloat[$p=0.1$]{
			\begin{tabular}{llll}
				\toprule
				Estimator	& CP & CIL & $I$  \\ 
				\hline
				Exact & 0.9776 & 0.4859 & 0.9281  \\ 
				Wald & 0.6250 & 0.2701 & 0.7477  \\ 
				Arc & 0.6110 & 0.2531 & 0.7396  \\ 
				Arc.CC & 0.9776 & 0.4757 & 0.9292  \\ 
				Pois & 0.5560 & 0.2630 & 0.6940  \\ 
				Wils & 0.9310 & 0.3661 & 0.9373  \\ 
				Wils.CC & 0.9860 & 0.4338 & 0.9321  \\ 
				BCG & 0.6110 & 0.2701 & 0.7369  \\ 
				AgreC & 0.9310 & 0.4015 & 0.9331  \\ 
				Ag.add4 & 0.9310 & 0.4042 & 0.9328  \\ 
				midP & 0.9860 & 0.3721 & 0.9395  \\ 
				\bottomrule
			\end{tabular}
		}\hfill
		\subfloat[$p=0.5$]{
			\begin{tabular}{llll}
				\toprule
				CP & CIL & $I$ \\ 
				\hline
				0.9810 & 0.6012 & 0.9149  \\ 
				0.8990 & 0.5877 & 0.8936  \\ 
				0.8990 & 0.5508 & 0.8975  \\ 
				0.9810 & 0.6286 & 0.9122  \\ 
				0.9880 & 0.7937 & 0.8948  \\ 
				0.9810 & 0.5076 & 0.9248  \\ 
				0.9810 & 0.5777 & 0.9174 \\ 
				0.9810 & 0.5877 & 0.9163 \\ 
				0.9810 & 0.5131 & 0.9242  \\ 
				0.9810 & 0.5106 & 0.9245  \\ 
				0.9810 & 0.5517 & 0.9201  \\ 
				
				\bottomrule
			\end{tabular}
		}\hfill
		\subfloat[$p=0.8$]{
			\begin{tabular}{llll}
				\toprule
				CP & CIL & $I$ \\ 
				\hline
				0.9943 & 0.5264 & 0.9196 \\ 
				0.8780 & 0.4305 & 0.8982\\
				0.8490 & 0.4034 & 0.8835  \\ 
				0.9943 & 0.5306 & 0.9191  \\ 
				1.0000 & 1.0311 & 0.8717  \\ 
				0.9660 & 0.4299 & 0.9371  \\ 
				0.9660 & 0.4987 & 0.9294 \\ 
				0.8780 & 0.4305 & 0.8982 \\ 
				0.9660 & 0.4544 & 0.9343  \\ 
				0.9660 & 0.4547 & 0.9343 \\ 
				0.9660 & 0.4536 & 0.9344 \\ 
				\bottomrule
			\end{tabular}
		}
	\end{adjustbox}	
	\label{Prop_Tab1}	
\end{table}

\begin{table}[htb!]
	\centering
	\caption{Confidence interval index for the binomial proportion using  $n=20$}
	\begin{adjustbox}{width=\textwidth}		
		\subfloat[$p=0.1$]{
			\begin{tabular}{llll}
				\toprule
				Estimator	& CP & CIL & $I$  \\ 
				\hline
				Exact & 0.9853 & 0.3123 & 0.9472\\ 
				Wald & 0.8850 & 0.2398 & 0.9270  \\ 
				Arc & 0.8730 & 0.2322 & 0.9210  \\ 
				Arc.CC & 0.9853 & 0.3041 & 0.9483  \\ 
				Pois & 0.8440 & 0.2313 & 0.9036  \\ 
				Wils & 0.9580 & 0.2664 & 0.9589  \\ 
				Wils.CC & 0.9870 & 0.2970 & 0.9489  \\ 
				BCG & 0.8440 & 0.2398 & 0.9024  \\ 
				AgreC & 0.9580 & 0.2883 & 0.9561  \\ 
				Ag.add4 & 0.9580 & 0.2897 & 0.9559  \\ 
				midP & 0.9870 & 0.2662 & 0.9530  \\ 
				\bottomrule
			\end{tabular}
		}\hfill
		\subfloat[$p=0.5$]{
			\begin{tabular}{llll}
				\toprule
				CP & CIL & $I$ \\ 
				\hline
				0.9670 & 0.4462 & 0.9350 \\ 
				0.9670 & 0.4270 & 0.9372  \\ 
				0.9670 & 0.4135 & 0.9388  \\ 
				0.9670 & 0.4582 & 0.9337  \\ 
				0.9940 & 0.5755 & 0.9144  \\ 
				0.9670 & 0.3928 & 0.9412  \\ 
				0.9670 & 0.4343 & 0.9364  \\ 
				0.9670 & 0.4270 & 0.9372 \\ 
				0.9670 & 0.3943 & 0.9410  \\ 
				0.9670 & 0.3930 & 0.9412  \\ 
				0.9670 & 0.4126 & 0.9389  \\ 
				
				\bottomrule
			\end{tabular}
		}\hfill
		\subfloat[$p=0.8$]{
			\begin{tabular}{llll}
				\toprule
				CP & CIL & $I$ \\ 
				\hline
				0.9780 & 0.3703 & 0.9415  \\ 
				0.9220 & 0.3351 & 0.9360  \\ 
				0.8960 & 0.3245 & 0.9221  \\ 
				0.9780 & 0.3725 & 0.9412  \\ 
				1.0000 & 0.7400 & 0.8965 \\ 
				0.9520 & 0.3261 & 0.9526  \\ 
				0.9910 & 0.3672 & 0.9390 \\ 
				0.9520 & 0.3351 & 0.9515  \\ 
				0.9520 & 0.3392 & 0.9510  \\ 
				0.9520 & 0.3391 & 0.9510  \\ 
				0.9520 & 0.3361 & 0.9514  \\ 
				
				\bottomrule
			\end{tabular}
		}
	\end{adjustbox}	
	\label{Prop_Tab2}	
\end{table}

% latex table generated in R 3.1.3 by xtable 1.7-4 package
% Mon Aug 15 15:08:07 2016
\begin{table}[htp!]
	\centering
	\caption{Confidence interval index for the binomial proportion using  $n=100$}
	\begin{adjustbox}{width=\textwidth}		
		\subfloat[$p=0.1$]{
			\begin{tabular}{llll}
				\toprule
				Estimator	& CP & CIL & $I$ \\ 
				\hline
				Exact & 0.9740 & 0.1798 & 0.9677  \\ 
				Wald & 0.8820 & 0.1599 & 0.9369  \\ 
				Arc & 0.9420 & 0.1579 & 0.9711  \\ 
				Arc.CC & 0.9560 & 0.1774 & 0.9715  \\ 
				Pois & 0.9420 & 0.1553 & 0.9715  \\ 
				Wils & 0.9740 & 0.1655 & 0.9697  \\ 
				Wils.CC & 0.9740 & 0.1838 & 0.9671  \\ 
				BCG & 0.9420 & 0.1599 & 0.9708  \\ 
				AgreC & 0.9740 & 0.1748 & 0.9684  \\ 
				Ag.add4 & 0.9740 & 0.1753 & 0.9683  \\ 
				midP & 0.9420 & 0.1646 & 0.9701 \\ 
				
				\bottomrule
			\end{tabular}
		}\hfill
		\subfloat[$p=0.5$]{
			\begin{tabular}{llll}
				\toprule
				CP & CIL & $I$  \\ 
				\hline
				0.9670 & 0.2869 & 0.9544  \\ 
				0.9390 & 0.2745 & 0.9534  \\ 
				0.9390 & 0.2710 & 0.9539  \\ 
				0.9670 & 0.2901 & 0.9540 \\ 
				0.9890 & 0.3732 & 0.9387  \\ 
				0.9390 & 0.2647 & 0.9547  \\ 
				0.9670 & 0.2832 & 0.9549 \\ 
				0.9670 & 0.2745 & 0.9560  \\ 
				0.9390 & 0.2648 & 0.9547  \\ 
				0.9390 & 0.2645 & 0.9547  \\ 
				0.9390 & 0.2702 & 0.9540  \\ 
				%0.9670 & 0.2770 & 0.9557 \\ 
				\bottomrule
			\end{tabular}
		}\hfill
		\subfloat[$p=0.8$]{
			\begin{tabular}{llll}
				\toprule
				CP & CIL & $I$ \\ 
				\hline
				0.9670 & 0.1638 & 0.9713  \\ 
				0.9230 & 0.1554 & 0.9609  \\ 
				0.9500 & 0.1544 & 0.9760 \\ 
				0.9500 & 0.1642 & 0.9746  \\ 
				1.0000 & 0.3412 & 0.9404  \\ 
				0.9320 & 0.1542 & 0.9661 \\ 
				0.9670 & 0.1638 & 0.9713  \\ 
				0.9320 & 0.1554 & 0.9659  \\ 
				0.9320 & 0.1558 & 0.9659  \\ 
				0.9490 & 0.1558 & 0.9752  \\ 
				0.9500 & 0.1548 & 0.9759  \\ 
				
				\bottomrule
			\end{tabular}
		}
	\end{adjustbox}	
	\label{Prop_Tab3}	
\end{table}

We find that the confidence interval length improves with increasing sample size. In addition, most of the empirical coverage probabilities of the estimators become much closer to the nominal level of 0.95 as the sample size increases. In particular, the Pois estimator for estimating $p=0.1,$ improves drastically from $\hat{\eta}=0.556$ to  $\hat{\eta}=0.942$ respectively for sample sizes 10 and 100. Together with the corresponding confidence interval lengths, the values of index for the Pois estimator increases from 0.6940 (compared with the best estimator's index of 0.9395) to 0.9715 (joint best with Arc.CC) for sample sizes 10 and 100 respectively. However, for other values of $p$ (more generally $p>0.1$), the Pois estimator overestimates the coverage probability and has larger confidence intervals relative to the other estimators. Therefore, the Pois estimator has smaller index values compared with the other estimators, and hence, is not appropriate for the estimation of $p.$  

Furthermore, the Exact estimator overestimates the coverage probability in all cases. In addition, it has large confidence interval lengths and these are shown in its index values. This is consistent with results reported in \cite{Pires2008}. In most cases, estimators such as Wilson, AgreC, Ag.add4 and midP have relatively good coverage properties and interval lengths and are shown in their $I$ values usually approaching 1.

In general, for the estimation of a proportion, the index is able to distinguish between estimators that are appropriate or not based on their interval lengths and coverage probabilities.

\section{Application}\label{sec4}

To illustrate the application of our index, we consider the paper by \cite{Gulhar2012}. The authors compared several confidence interval estimators for the coefficient of variation (CV). The coefficient of variation is defined as the variability of a random variable relative to its mean. It is usually expressed as a percentage. The confidence interval estimators of the CV were compared based on their interval lengths and the empirical coverage probabilities. The authors used separate plots for the coverage probabilities and the interval lengths across different sample sizes, CV values and distributions. We take a different approach in this paper by constructing plots showing simultaneously the coverage probabilities and the confidence interval index, $I.$

\begin{table}[htp!]
	\caption{List of confidence interval estimators in \cite{Gulhar2012}}
	\begin{adjustbox}{width=\textwidth}
		%	\large{
		\begin{tabular}{ll}
			\hline
			
			Abbreviation &  Confidence Interval type\\
			\hline
			NP.BS        & Bootstrap Percentile \\
			PBS          & Bootstrap-$t$   \\
			Mill         &  \cite{Miller1991} interval from asymptotic normal approximation  \\
			BSMill       & Modified median Miller estimator based on critical values from bootstrap samples\\ 
			BS C.P        & \cite{Curto2009} modified median  estimator based on BS sample\\
			
			S.K        & \cite{Sharma1994} interval from inverted CV  \\
			C.P         &  \cite{Curto2009} \textit{iid} assumption interval\\
			McK          &  \cite{McKay1932} Interval from chi-square approximation  \\
			MMcK         & Modified McKay's interval \cite{Vangel2012}\\
			Panich       & \cite{Panich2009} modified McKay's interval \\
			Prop      & \cite{Gulhar2012} interval from chi-square approximation \\	
			MedMill     & Median Modified Miller Estimator \\
			MedMcK     & Median Modified of McKay interval \\
			MedMMcK     & Median Modification of Modified McKay's interval\\
			Med C.P     & Median Modified  \cite{Curto2009} interval\\
			\hline 
		\end{tabular}
	\end{adjustbox} 	
	
	\label{Abbrev}%}
\end{table}

\begin{figure}[htp!]
	\centering
	\subfloat[$n=15,~ CV=0.1$]{%
		\includegraphics[height=4.5cm,width=.33\textwidth]{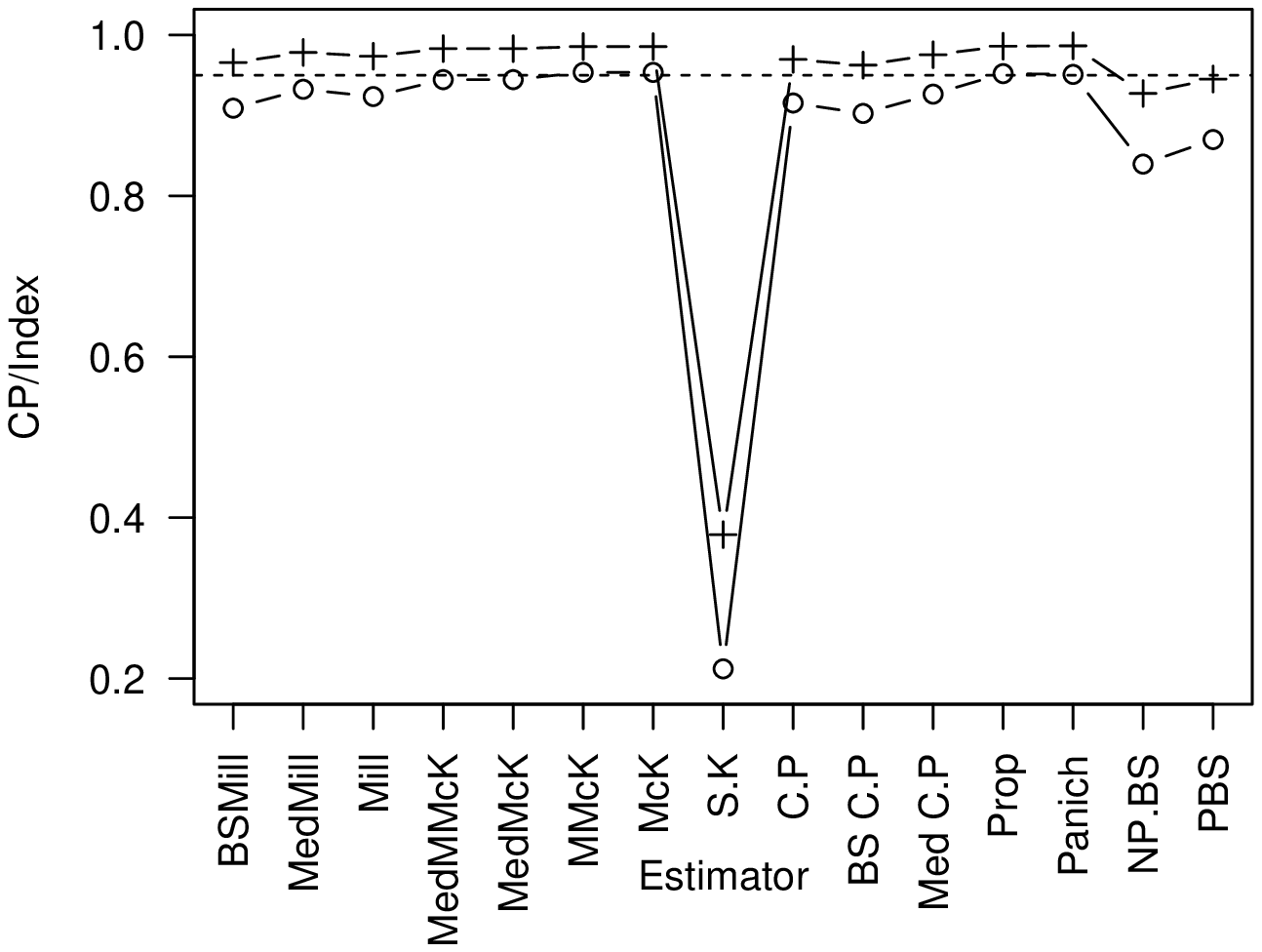}}\hfill
	\subfloat[$n=15,~ CV=0.3$]{%
		\includegraphics[height=4.5cm,width=.33\textwidth]{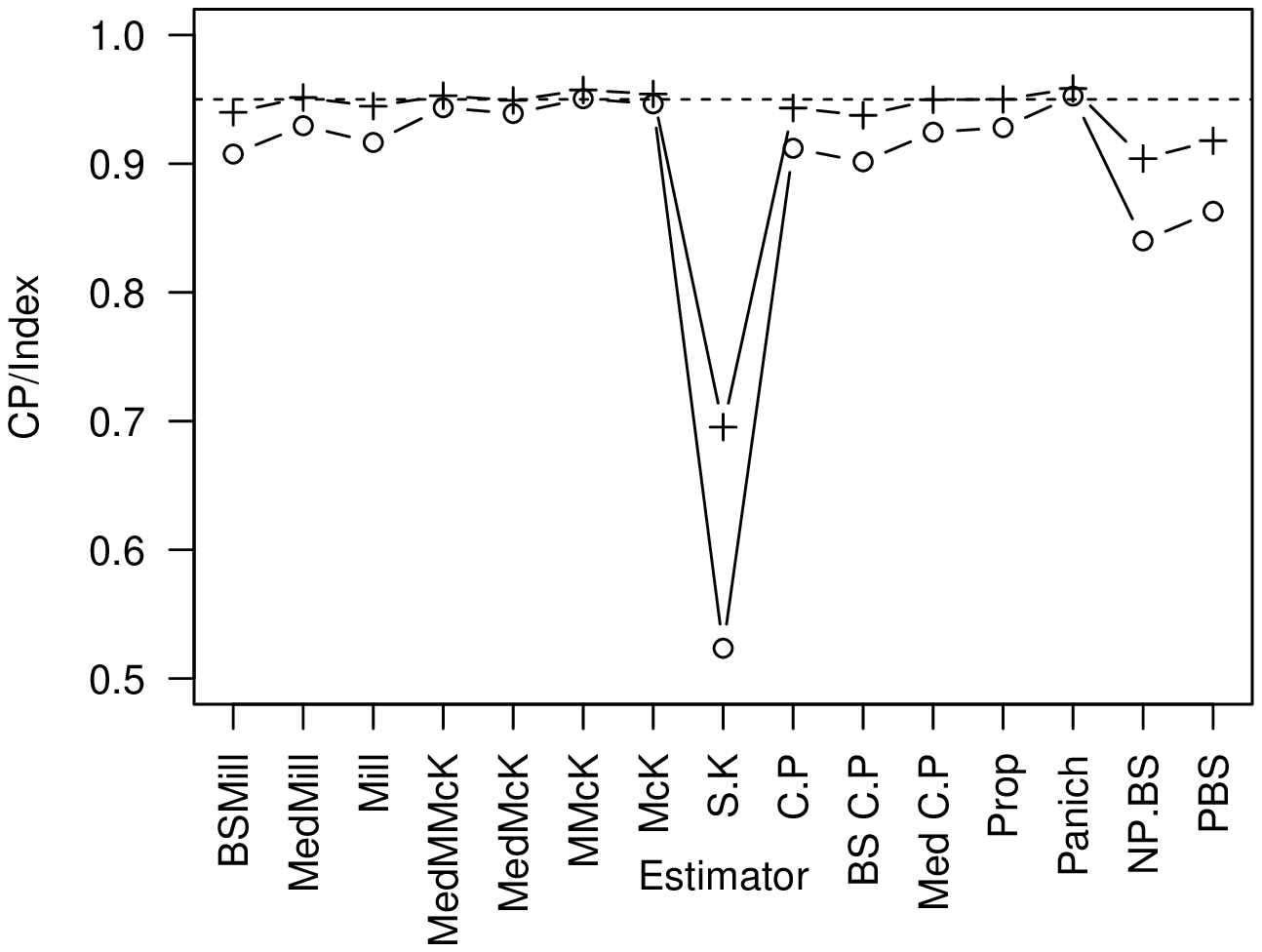}}\hfill
	\subfloat[$n=15,~ CV=0.5$]{%
		\includegraphics[height=4.5cm,width=.33\textwidth]{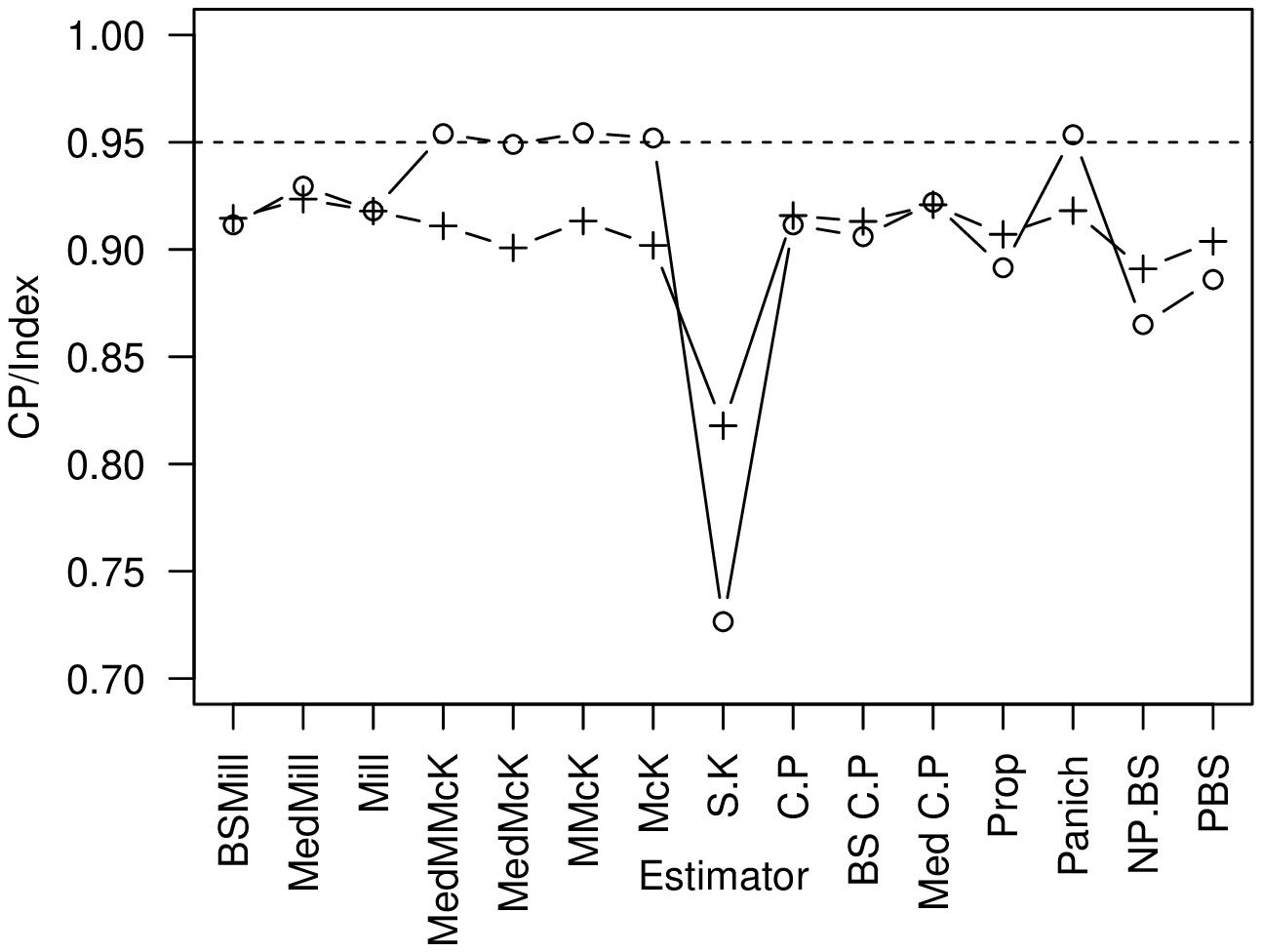}}\\
	\subfloat[$n=25,~ CV=0.1$]{%
		\includegraphics[height=4.5cm,width=.33\textwidth]{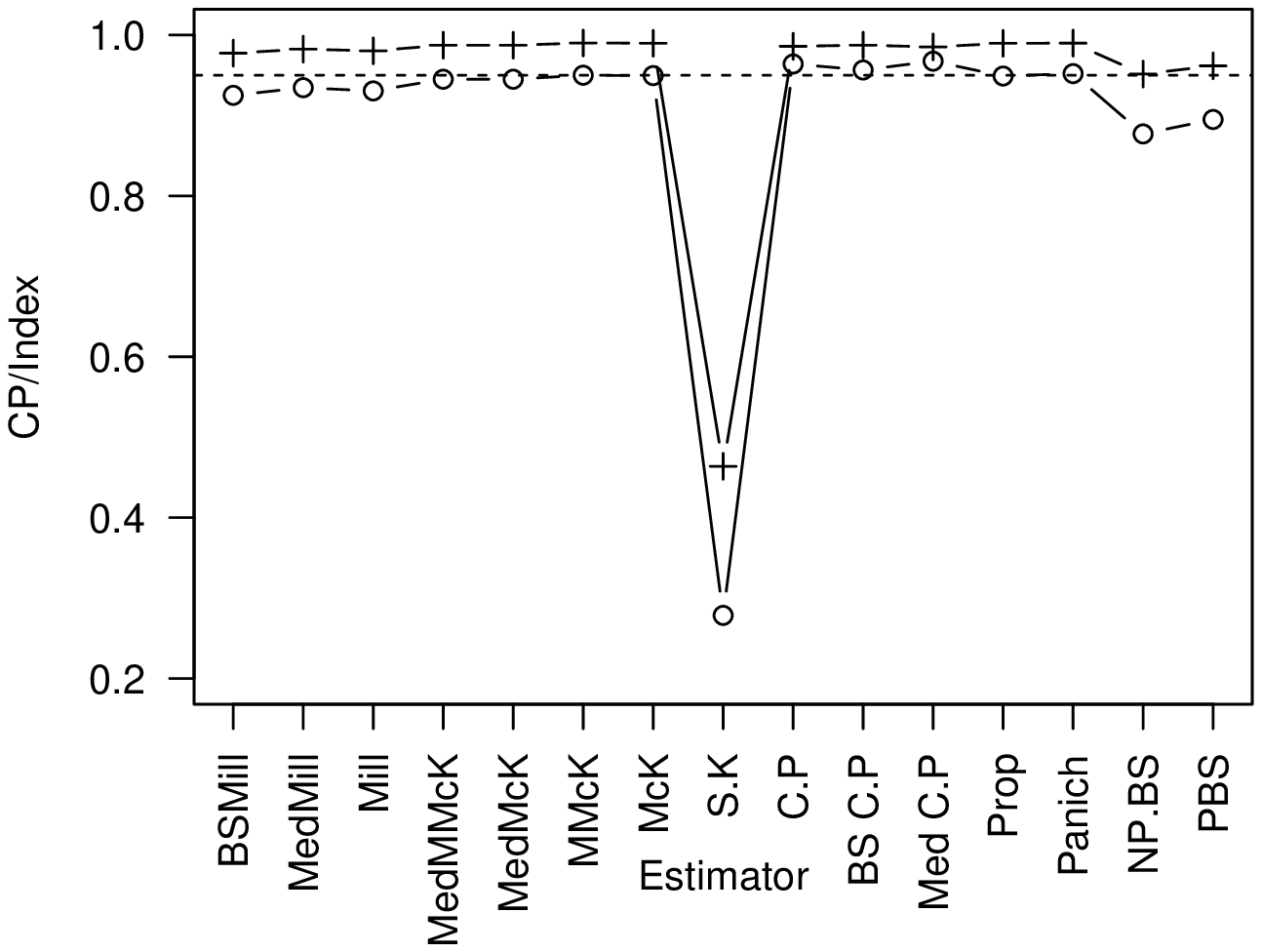}}\hfill
	\subfloat[$n=25,~ CV=0.3$]{%
		\includegraphics[height=4.5cm,width=.33\textwidth]{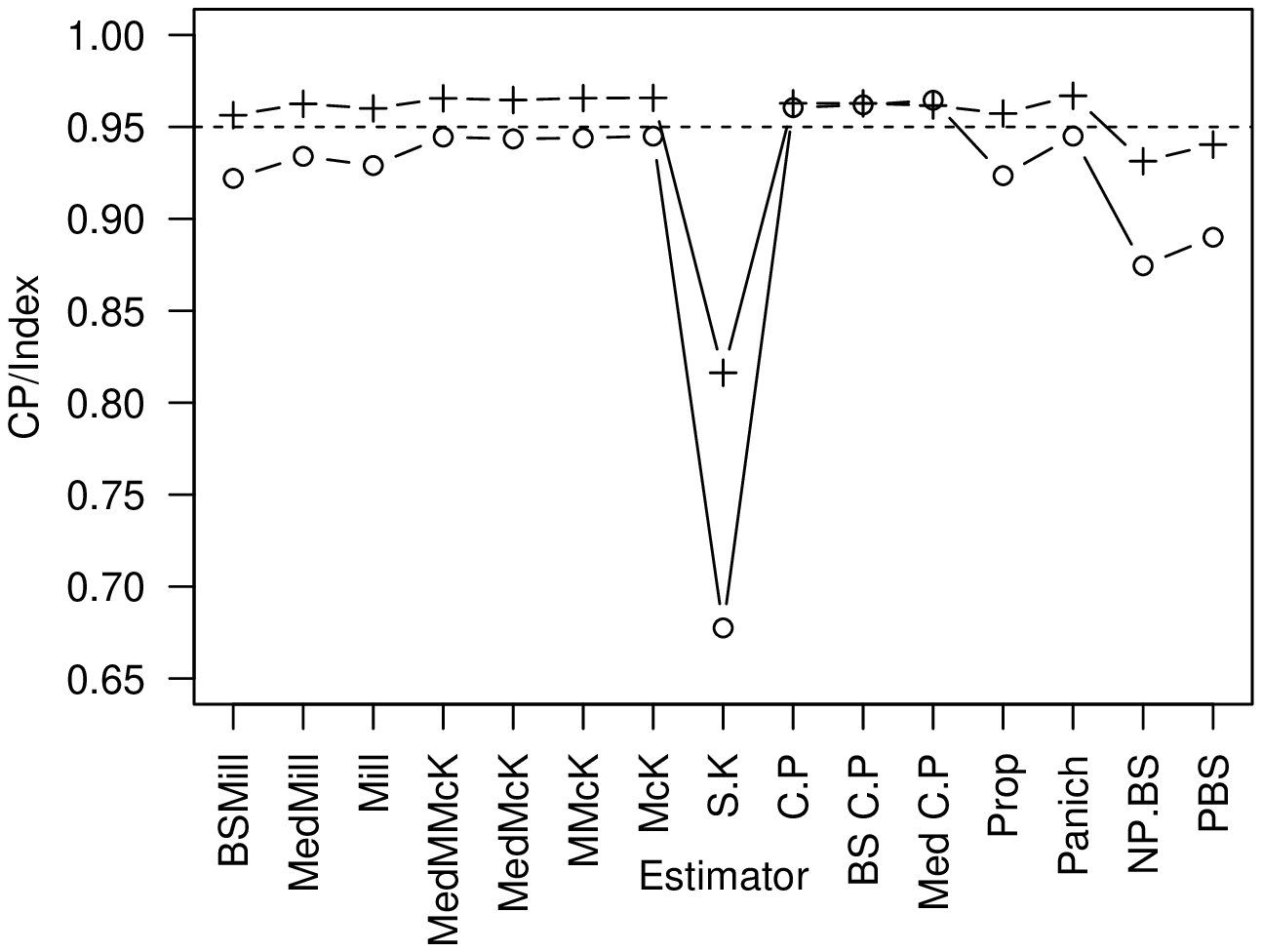}}\hfill
	\subfloat[$n=25,~ CV=0.5$]{%
		\includegraphics[height=4.5cm,width=.33\textwidth]{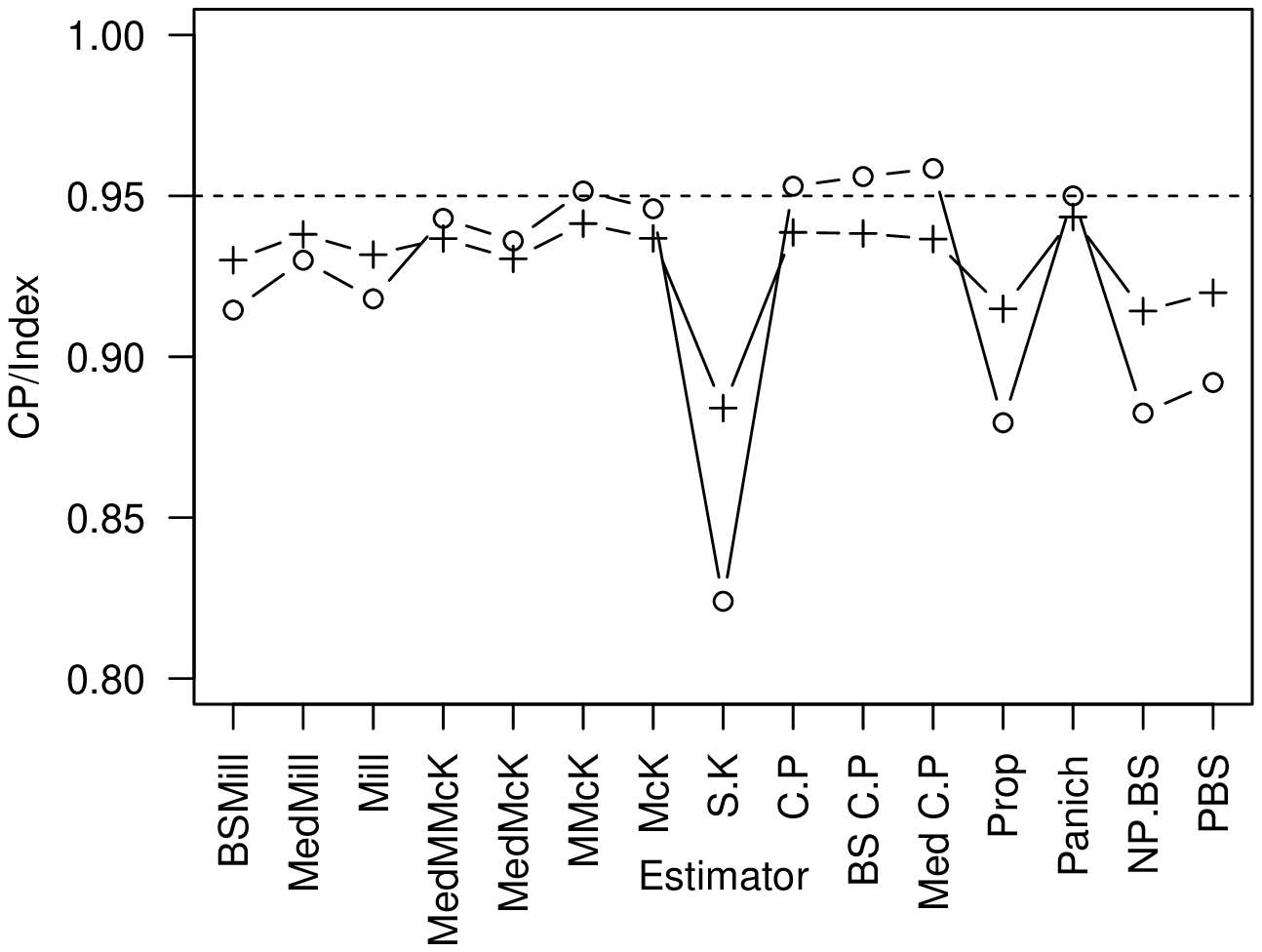}}\\
	\subfloat[$n=50,~ CV=0.1$]{%
		\includegraphics[height=4.5cm,width=.33\textwidth]{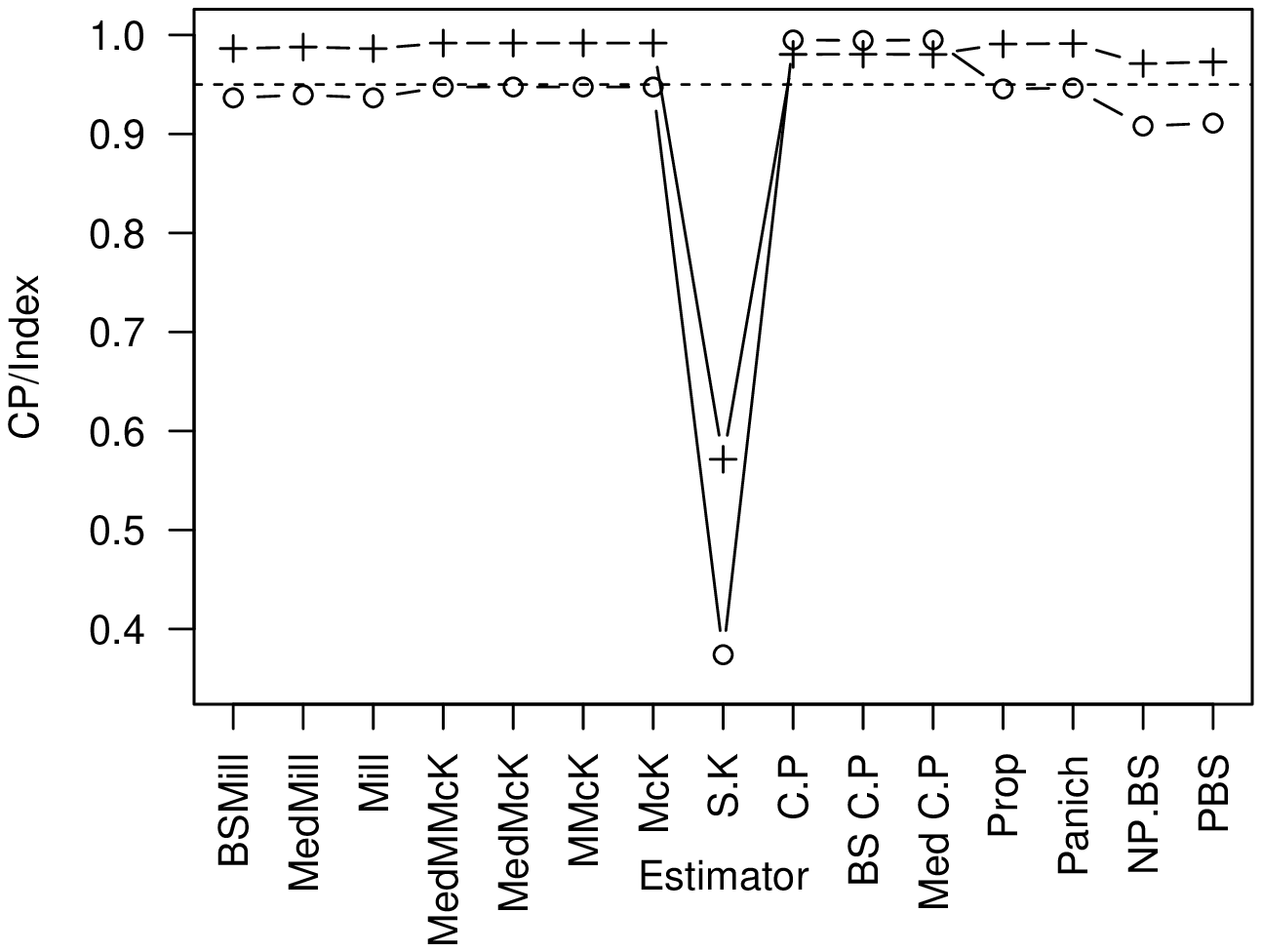}}\hfill
	\subfloat[$n=50,~ CV=0.3$]{%
		\includegraphics[height=4.5cm,width=.33\textwidth]{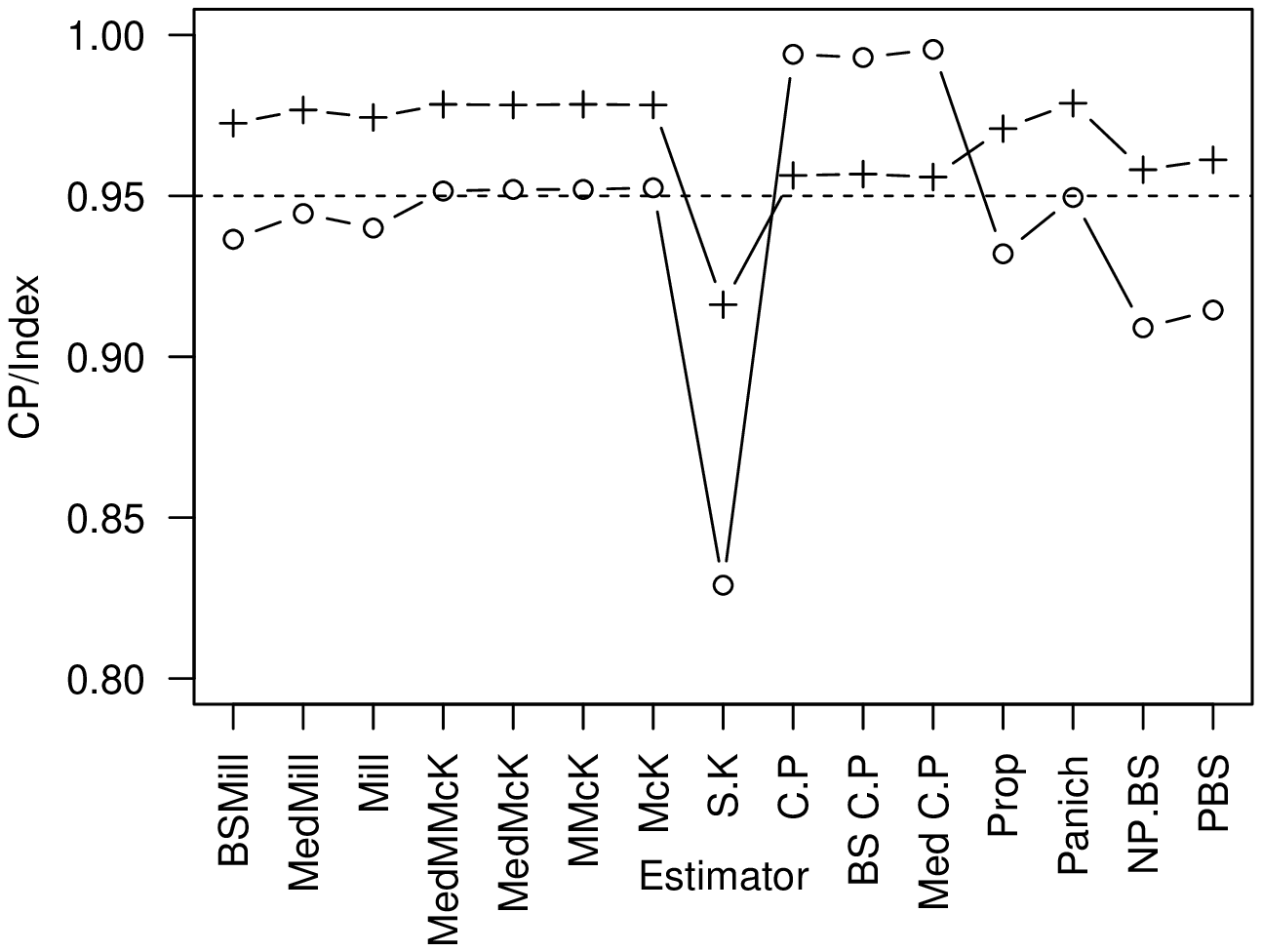}}\hfill
	\subfloat[$n=50,~ CV=0.5$]{%
		\includegraphics[height=4.5cm,width=.33\textwidth]{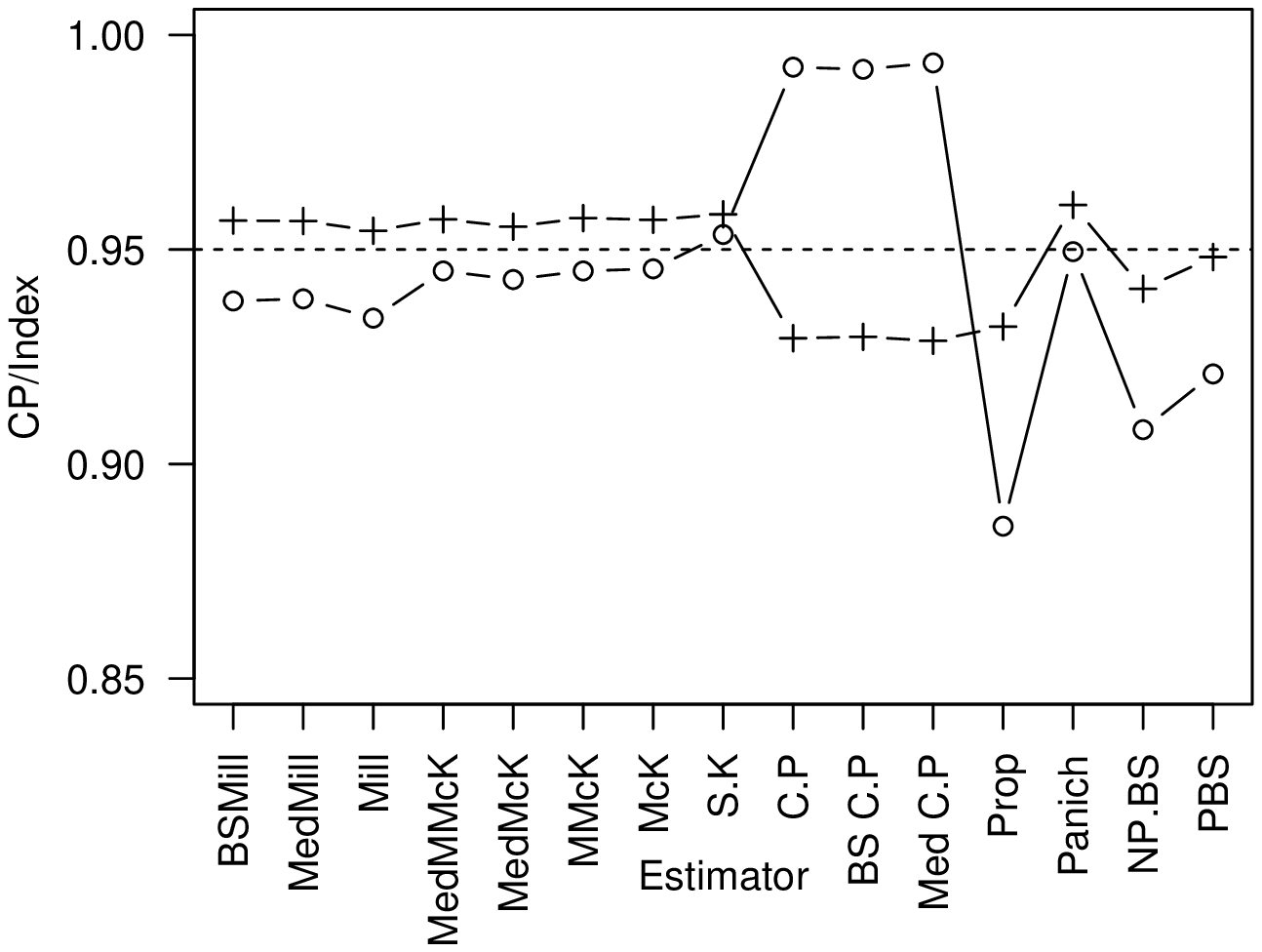}}\\
	\subfloat[$n=100,~ CV=0.1$]{%
		\includegraphics[height=4.5cm,width=.33\textwidth]{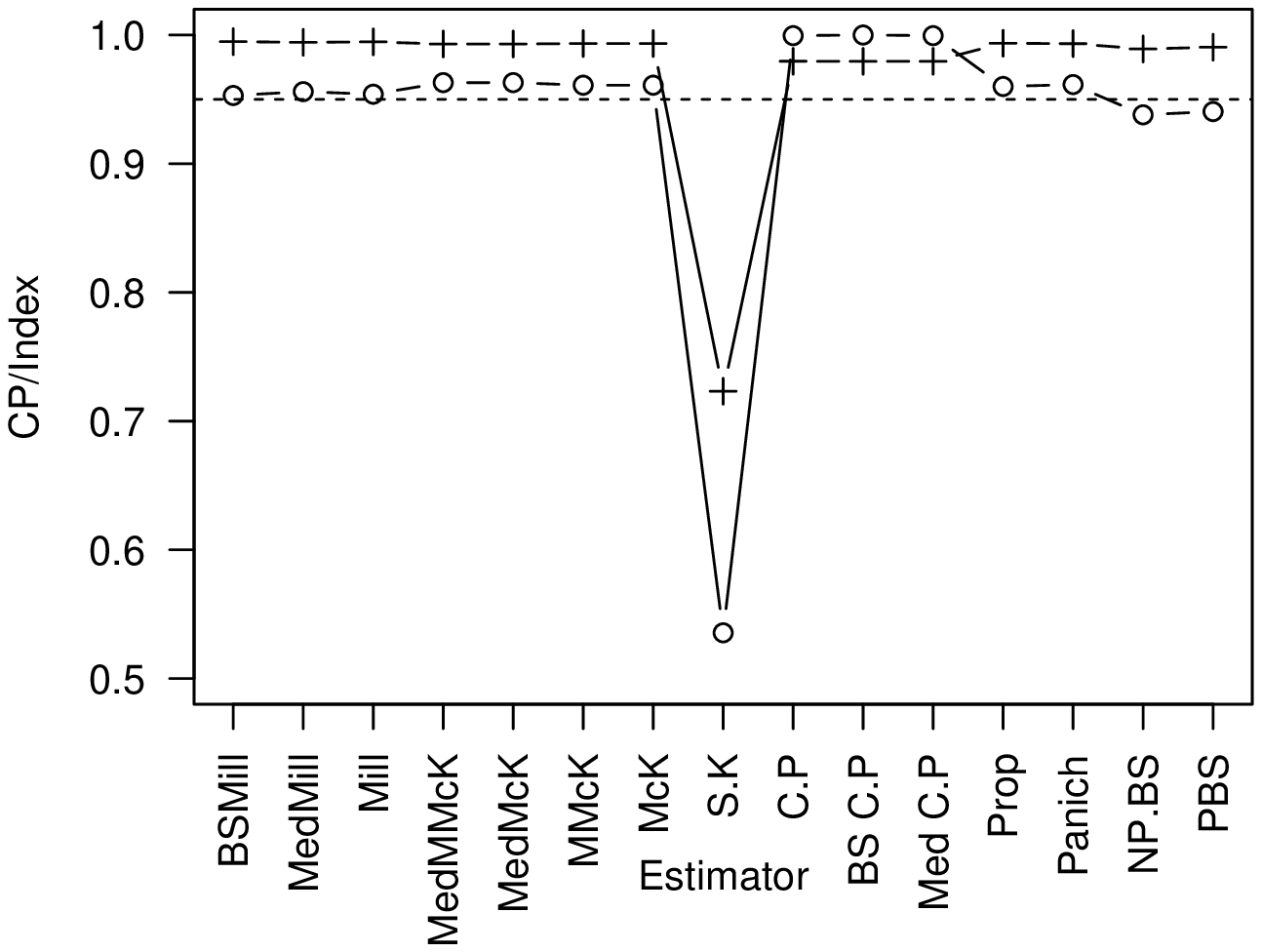}}\hfill
	\subfloat[$n=100, CV=0.3$]{%
		\includegraphics[height=4.5cm,width=.33\textwidth]{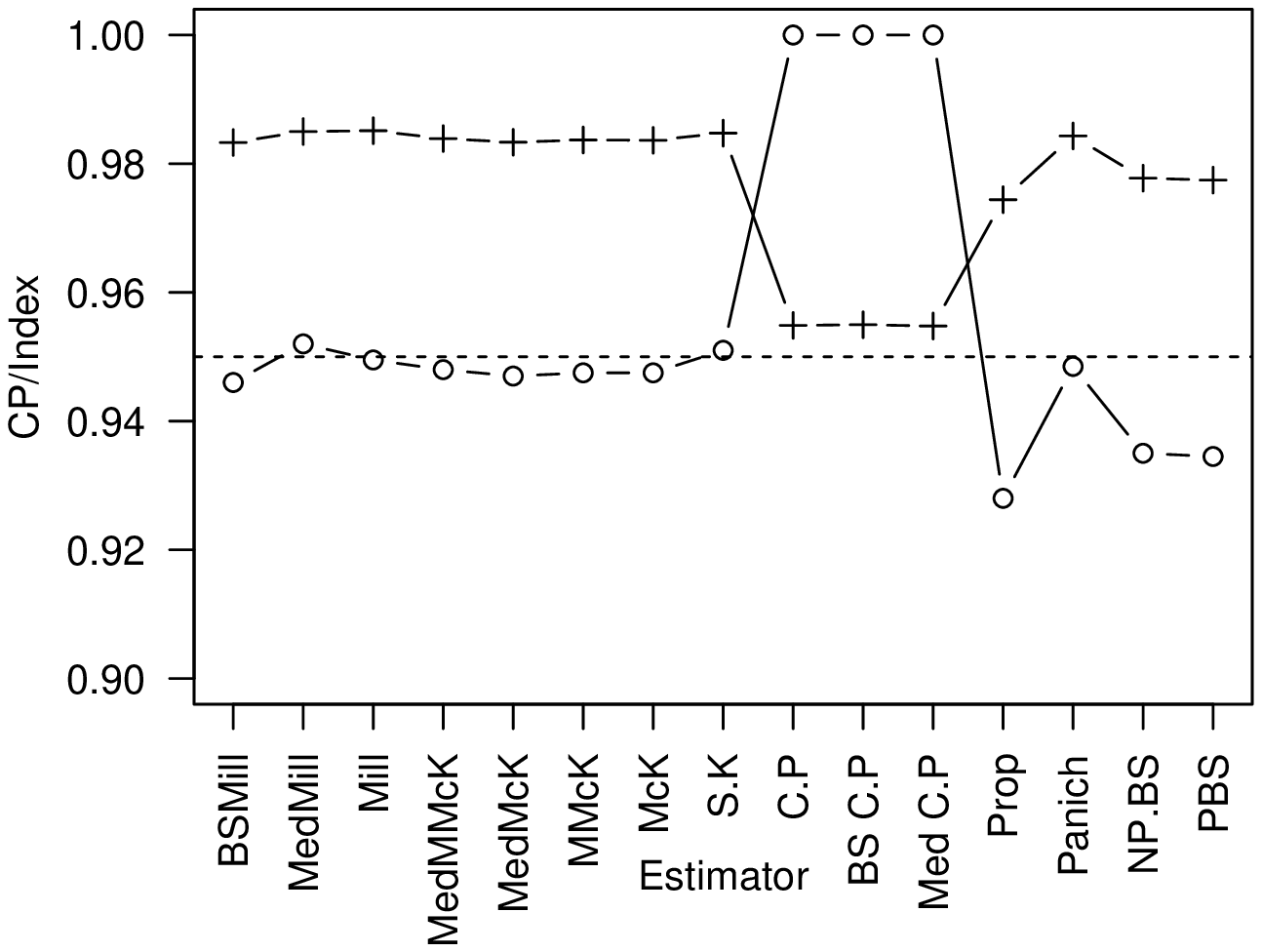}}\hfill
	\subfloat[$n=100,~ CV=0.5$]{%
		\includegraphics[height=4.5cm,width=.33\textwidth]{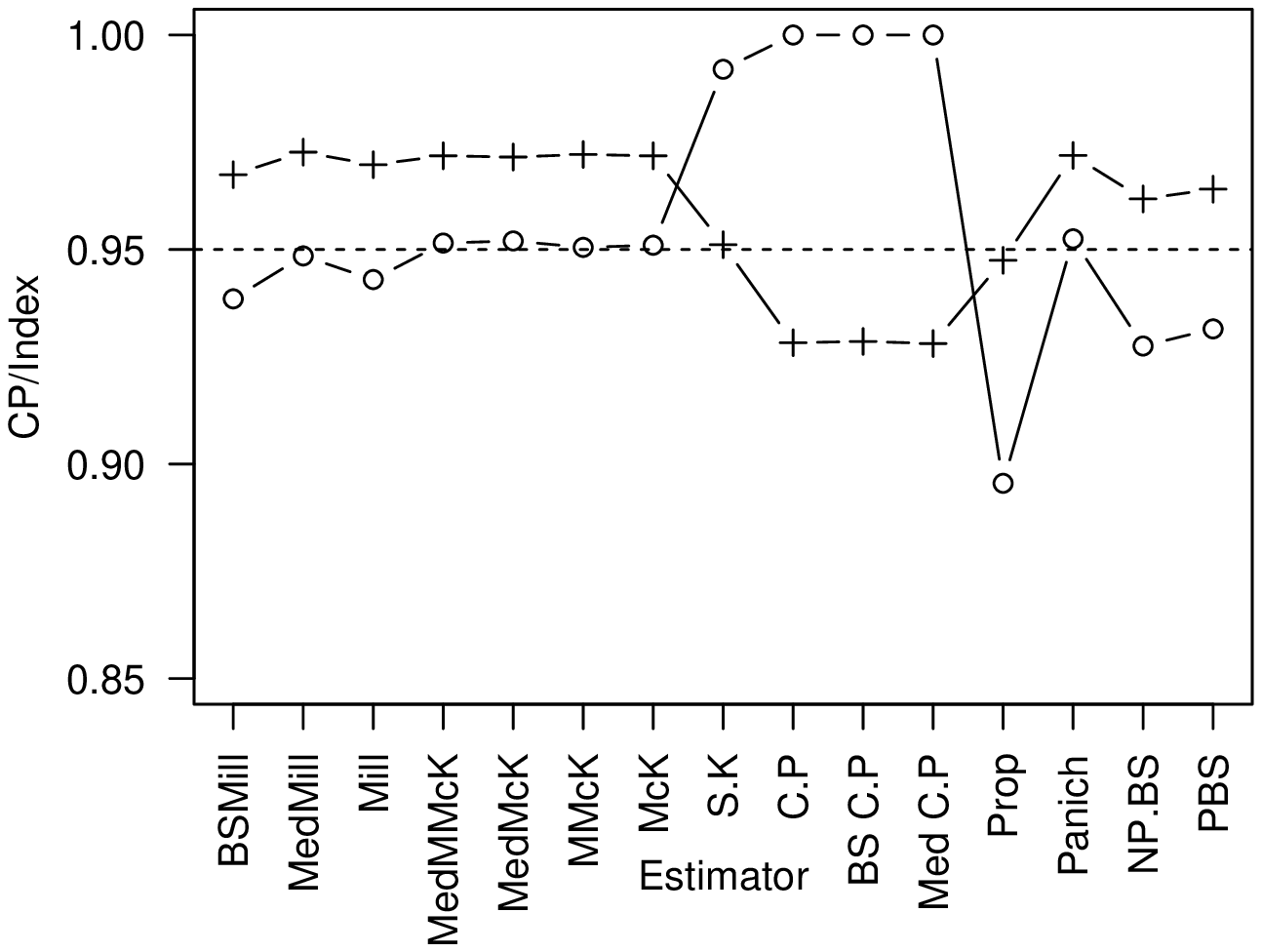}}\\
	\caption{Line plot of coverage probability of estimators with confidence interval index superimposed.   $\circ :$ empirical coverage probability. $+ : ~ I$ value. - - - : nominal coverage.}
	\label{fig2}
\end{figure}

The various confidence intervals considered and their abbreviations are presented in Table \ref{Abbrev}. We compute the confidence interval indexes for the estimators in Table 8 using the values in Table 4 of \cite[page 63]{Gulhar2012}. In addition, we assess the conclusions reached in that paper with that of the computed confidence interval indexes. The confidence interval indexes for each combination of $n$ and CV are shown in Tables \ref{A1}-\ref{A4} in Appendix B. In addition, the plots of the coverage probabilities and the corresponding confidence interval index values are presented in Figure \ref{fig2}. We can now make inferences from the graphs and compare these with the conclusions reached in \cite{Gulhar2012}.

Firstly, it can be seen that the most visible estimator that performs badly is the S.K estimator. It has mostly low coverage probability and this is reflected in it having smaller values on the index. It must be noted that some of the corresponding interval lengths of the S.K estimator were 2 to 8 times shorter than the other interval lengths. However, having a shorter interval length with low coverage probability is not practically desirable. As the sample size increases, there is a remarkable increase in the performance of the S.K estimator especially for $CV=0.5.$ Therefore, we can conclude that, in this case, the index discriminates the bad estimator from the good ones even though shorter interval lengths were recorded. 

Secondly, \cite[page 57]{Gulhar2012} concludes that ``By $n = 100,$ almost all intervals are performing at a similar level (Figure 1). All C.P intervals (C.P, Med C.P, and BS C.P) over exceeded the expected coverage probability of 95\% and reached 100\% and are clear outliers". From Figure \ref{fig2}, it can easily be seen from  the bottom panel (i.e. for $n=100$) that the index values for these estimators are smaller compared to the other estimators: this indicates that the C.P-based estimators are inappropriate for the estimation of the CV relative to the other estimators.

Thirdly, we can plot the index against sample size and CV values as an alternative to the four cases: CP against sample size; CP against CV; Interval length against $n;$ and interval length against CV. These graphs, not shown here, lead to the same conclusions obtained in \cite{Gulhar2012}. 

In general, the index values are consistent with the conclusions from the CP and the interval lengths. Therefore, the index provides a useful, but computationally inexpensive method for measuring the relative performance of the estimators of confidence intervals for CV.  

\section{Conclusion}\label{sec5}

In this paper, an index for measuring the performance of confidence interval estimators was proposed. The index is based on the traditional trade-off between confidence interval length and empirical coverage probability. Unlike the confidence interval length which has range, $\R^+$,  the index has range of values within that of the coverage probability. We showed that index values close to 1 indicate a good confidence interval estimator whereas values far removed from 1 indicate a bad confidence interval estimator. Thus, it can easily be superimposed on a plot of coverage probabilities to aid in the selection of estimators with good coverage probabilities and interval lengths. The index can be used alone or to complement the coverage probability for measuring the performance of confidence interval estimators. 

In all the simulations and practical application, we assessed the performance of estimators through the sizes of the values of the index. However, an issue of practical importance is the statistical difference between indexes. In practice, we propose that a hypothesis of equality or otherwise can be performed on any observed differences between indexes. Since the sampling distribution of the index remains an open problem, a non-parametric or the estimation of standard errors based on resampling methods can be used in such a test.  

%\begin{acknowledgements}
%\end{acknowledgements}

%\bibliography{Index}   % name your BibTeX data base

\begin{thebibliography}{99}
	
	%\bigskip
	%\baselineskip=18pt
	%\def\ref{\noindent\hangindent 25pt}
	%\ref Aitchison, J. (1964). Bayesian tolerance regions. {\it Jou.
	%of Royal Statistical Society}, B, {\bf 127}, 161-175.
	
	
	\bibitem{Agresti2000}
	Agresti, A., Caffo, B.: {Simple and Effective Confidence Intervals for
		Proportions and Differences of Proportions Result from Adding Two Successes
		and Two Failures}.
	\newblock \emph{ The American Statistician} 54(4), 280--288 (2000)
	
	\bibitem{Agresti1998}
	Agresti, A., Coull, B.A.: {Approximate is Better than ``Exact" for Interval
		Estimation of Binomial Proportions}.
	\newblock \emph{The American Statistician} 52(2), 119--126 (1998)
	
	\bibitem{Banik2010}
	Banik, S., Kibria, B.M.G.: {Comparison of Some Parametric and Nonparametric Type One Sample Confidence Intervals for Estimating the Mean of a Positively Skewed Distribution.}
	\newblock \emph{Communications in Statistics -Simulation and Computation} 39,
	361--389 (2010)
	
	\bibitem{Beran1987}
	Beran, R. (1987). {Prepivoting to Reduce Level Error of Confidence Sets}.
	\newblock \emph{Biometrika} 74(3), 457--468 
	
	\bibitem{Brown2001}
	Brown, L.D., Cai, T.T., DasGupta, A. (2001). {Interval Estimation for a Binomial
		Proportion}. 
	\newblock \emph{Statistical Science} 16(2), 101--117 
	
	\bibitem{Brown2002}
	Brown, L.D., Cai, T.T., DasGupta, A. (2002). {Confidence Intervals for a Binomial
		Proportion and Asymptotic Expansions}.
	\newblock \emph{The Annals of Statistics} 30(1), 160--201 
	
	\bibitem{Curto2009}
	Curto, J.D., Pinto, J.C. (2009). {The Coefficient of Variation Asymptotic Distribution
		in the Case of Non-iid Random Variables}.
	\newblock Journal of Applied Statistics 36(1), 21--32 
	
	\bibitem{Efron1993}
	Efron, B., Tibshirani, R.J. (1993). {An Introduction to the Bootstrap}.
	\newblock Chapman and Hall, London 
	
	\bibitem{Gulhar2012}
	Gulhar, M., {Golam Kibria}, B.M., Albatineh, A.N., Ahmed, N.U. (2012). {A Comparison of Some Confidence Intervals for Estimating the population coefficient of variation: A simulation study}.
	\newblock \emph{SORT} 36(1), 45--68 
	
	\bibitem{Huber1992}
	Huber, P.J. (1992). Robust estimation of a location parameter.
	\newblock In: Breakthroughs in Statistics, pp. 492--518. Springer 
	
	\bibitem{Johnson1978}
	Johnson, N.J. (1978). {Modified t Tests and Confidence Intervals for Asymmetrical
		Populations Modified L Tests and Confidence Intervals for Asymmetrical
		Populations}.
	\emph{Journal of the American Statistical Association} 73(363),
	536--544 
	
	\bibitem{Kilian2000}
	Kilian, L., Chang, P.L. (2000). {How accurate are confidence intervals for impulse
	responses in large var models?}
	\newblock \emph{Economics Letters} 69(3), 299--307 
	
	\bibitem{Lee2003}
	Lee, S.M.S., Young, G.A. (2003). {Prepivoting by Weighted Bootstrap Iteration}.
	\newblock \emph{Biometrika} 90, 393--410 
	
	\bibitem{Leemis1996}
	Leemis, L., Trivedi, K. (1996). {A Comparison of Approximate Interval Estimators for
		the Bernoulli Parameter}.
	\newblock \emph{The American Statistician} 50(1), 1--20 
	
	\bibitem{Loh1987}
	Loh, W.Y. (1987). {Calibrating Confidence Coefficients}.
	\newblock \emph{Journal of the American Statistical Association} 82(397),
	155--162 
	
	\bibitem{Loh1988}
	Loh, W.Y. (1988). {Discussion: Theoretical Comparison of Bootstrap Confidence
		Intervals}.
	\newblock \emph{The Annals of Statistics} 16(3), 972--976 
	
	\bibitem{Loh1991}
	Loh, W.Y.: {Bootstrap Calibration for Confidence Interval Construction and
		Selection}.
	\newblock \emph{Statistica Sinica} 1(2), 477--491 (1991)
	
	\bibitem{Martin1990}
	Martin, M. (1990). {On the Double Bootstrap}.
	\newblock Tech. rep., (Report No. 347) Department of Statistics, Stanford
	University, California. 
	
	\bibitem{McKay1932}
	McKay, A.T.: {Distribution of the Coefficient of Variation and the Extended \emph{t}	Distribution}.
	\newblock \emph{Journal of the Royal Statistical Society} 95 (4), 695--698
	(1932)
	
	\bibitem{Miller1991}
	Miller, G.E. (1991). {Asymptotic Test Statistics for Coefficients of Variation}.
	\newblock\emph{ Communications in Statistics - Theory and Methods} 20(10),
	3351--3363 
	
	\bibitem{Nankervis2005}
	Nankervis, J.C. (2005). {Computational Algorithms for Double Bootstrap Confidence
		Intervals}.
	\newblock\emph{ Computational Statistics {\&} Data Analysis} 49, 461--475
	
	
	\bibitem{Panich2009}
	Panichkitkosolkul, W. (2009). {Improved Confidence Intervals for a Coefficient of Variation of a Normal Distribution}.
	\newblock \emph{Thailand Statistician} 7(2), 193--199 
	
	\bibitem{Pires2008}
	Pires, A.M., Amado, C. (2008). {Interval Estimators for a Binomial Proportion: Comparison of Twenty Methods}.
	\newblock \emph{REVSTAT} 6(2), 165--197 
	
	\bibitem{Sharma1994}
	Sharma, K., Krishna, H. (1994). {Asymptotic Sampling Distribution of Inverse
		Coefficient-of-Variation and its Applications}.
	\newblock \emph{IEEE Transactions on Reliability} 43(4), 630--633 
	
	\bibitem{Vangel2012}
	Vangel, M.G. (1996). {Confidence Intervals for a Normal Coefficient of Variation}.
	\newblock \emph{The American Statistician} 50(1), 21--26 
	
	\bibitem{Wilson1927}
	Wilson, E.B. (1927). {Probable Inference, the Law of Succession, and Statistical
		Inference}.
	\newblock \emph{Journal of the American Statistical Association} 22(158),
	209--212 
	
	\bibitem{Zaane2012}
	Zaane, B.V., Vergouwe, Y., Donders, A.R.T., Moons, K.G.M. (2012) {Comparison of
		Approaches to Estimate Confidence Intervals of Post-test Probabilities of
		Diagnostic Test Results in a Nested Case-Control Study}.
	\newblock \emph{ BMC Medical Research Methodology} \textbf{12}(166), 1--9 
	
\end{thebibliography}
%\bibliographystyle{unsrt}
%\bibliographystyle{plainnat}

%\begin{thebibliography}{}

%\end{thebibliography}

\appendix
%\begin{appendices}
\addtocontents{toc}{\protect\setcounter{tocdepth}{0}}
%\section*{Appendices}

%\section{Appendix}
%\section{\\Appendix} \label{App:AppendixA}
% the \\ insures the section title is centered below the phrase: AppendixA
%\textbf{Appendix A}

\section{Appendix A}

\begin{table}[H]%[htb!]
	\centering
	\caption{Summary statistics for the Confidence Interval Index of the Mean from  $Lognormal(0,1)$}
	\begin{tabular}{llcccc}
		\toprule
		&  & \multicolumn{4}{c}{$I$}  \\ \cmidrule{3-6} 
		$n$	&	Basic Statistics & Normal Theory & Johnson $t$ &Bootstrap Percentile  & BCa\\
		\hline
		\multirow{3}*{$10$}
		&	Mean & 0.6574 & 0.6719 & 0.6547 & 0.6652 \\ 
		%&	Median & 0.6569 & 0.6717 & 0.6545 & 0.6649 \\ 
		&	Skewness & -0.0870 & -0.0719 & -0.0700 & -0.1204 \\ 
		&	Kurtosis & -0.0205 & 0.2760 & 0.1860 & -0.0512 \\ 
		&	St. dev & 0.0111 & 0.0104 & 0.0115 & 0.0108 \\  
		\hline\multirow{3}*{$50$}
		&	Mean & 0.7645 & 0.7669 & 0.7640 & 0.7645 \\ 
		%&	Median & 0.7649 & 0.7673 & 0.7644 & 0.7649 \\ 
		&Skewness & -0.2761 & -0.3045 & -0.2162 & -0.1518 \\ 
		&Kurtosis & -0.0811 & -0.0616 & -0.2474 & -0.1753 \\ 
		&St. dev & 0.0092 & 0.0090 & 0.0093 & 0.0089 \\  
		\hline\multirow{3}*{$100$}
		&	Mean & 0.8033 & 0.8048 & 0.8018 & 0.8005 \\ 
		%	&Median & 0.8033 & 0.8046 & 0.8021 & 0.8010 \\ 
		&Skewness & -0.1532 & -0.1717 & -0.1574 & -0.1004 \\ 
		&Kurtosis & -0.1021 & -0.1106 & -0.0155 & -0.1015 \\ 
		&St. dev & 0.0083 & 0.0081 & 0.0080 & 0.0083 \\
		\hline\multirow{3}*{$500$}
		&	Mean & 0.8806 & 0.8810 & 0.8780 & 0.8757 \\ 
		%&Median & 0.8816 & 0.8822 & 0.8786 & 0.8761 \\ 
		&Skewness & -0.6314 & -0.7137 & -0.2949 & -0.4518 \\ 
		&Kurtosis & -0.3605 & -0.1775 & -0.6620 & -0.0593 \\ 
		&St. dev & 0.0055 & 0.0053 & 0.0065 & 0.0065 \\ 
		\hline\multirow{3}*{$1000$}
		&	Mean & 0.9064 & 0.9066 & 0.9038 & 0.9024 \\ 
		%&Median & 0.9077 & 0.9079 & 0.9047 & 0.9033 \\ 
		&Skewness & -1.0806 & -1.1117 & -0.6161 & -0.5282 \\ 
		&Kurtosis & 0.6637 & 0.7065 & -0.3380 & -0.4309 \\ 
		&St. dev & 0.0045 & 0.0043 & 0.0061 & 0.0058 \\ 		
		
		\bottomrule
	\end{tabular}
	\label{skew2}
\end{table}

\begin{table}[htp]
	\centering
	\caption{Summary statistics for the Confidence Interval Index of the Mean from $Lognormal(0,0.2)$}
	\begin{tabular}{llcccc}
		\toprule
		&  & \multicolumn{4}{c}{$I$}  \\ \cmidrule{3-6} 
		$n$	&	Basic Statistics & Normal Theory & Johnson $t$ &Bootstrap Percentile  & BCa\\
		\hline
		\multirow{3}*{$10$}
		&	Mean & 0.9426 & 0.9549 & 0.9346 & 0.9332 \\ 
		%&Median & 0.9424 & 0.9558 & 0.9345 & 0.9332 \\ 
		&Skewness & -0.0624 & -1.1730 & 0.0143 & -0.0753 \\ 
		&Kurtosis & -0.0272 & 1.1878 & -0.0062 & -0.0660 \\ 
		&St. dev & 0.0050 & 0.0030 & 0.0054 & 0.0055 \\ 
		\hline\multirow{3}*{$50$}
		&		Mean & 0.9777 & 0.9793 & 0.9761 & 0.9759 \\ 
		%&Median & 0.9782 & 0.9801 & 0.9767 & 0.9761 \\ 
		&Skewness & -0.8278 & -1.7621 & -0.5530 & -0.4509 \\ 
		&Kurtosis & 0.2735 & 3.6100 & -0.0649 & -0.2337 \\ 
		&St. dev & 0.0034 & 0.0023 & 0.0039 & 0.0040 \\ 
		\hline\multirow{3}*{$100$}
		&	Mean & 0.9842 & 0.9849 & 0.9833 & 0.9831 \\ 
		%&Median & 0.9849 & 0.9855 & 0.9839 & 0.9834 \\ 
		&Skewness & -1.3544 & -1.7882 & -0.8309 & -0.8662 \\ 
		&Kurtosis & 2.1611 & 4.0482 & 0.3911 & 0.6067 \\ 
		&St. dev & 0.0026 & 0.0019 & 0.0030 & 0.0031 \\
		\hline\multirow{3}*{$500$}
		&	Mean & 0.9919 & 0.9920 & 0.9916 & 0.9915 \\ 
		% &Median & 0.9927 & 0.9926 & 0.9925 & 0.9925 \\ 
		&Skewness & -1.6490 & -1.7769 & -1.4613 & -1.4610 \\ 
		& Kurtosis & 2.8838 & 3.5023 & 1.9419 & 2.5344 \\ 
		& St. dev & 0.0021 & 0.0021 & 0.0024 & 0.0025 \\ 
		\hline\multirow{3}*{$1000$}
		&	Mean & 0.9938 & 0.9938 & 0.9934 & 0.9933 \\ 
		%&Median & 0.9946 & 0.9945 & 0.9942 & 0.9942 \\ 
		&Skewness & -1.7466 & -1.7604 & -1.5275 & -1.4532 \\ 
		&Kurtosis & 3.3484 & 3.3595 & 2.3416 & 1.8059 \\ 
		&St. dev & 0.0020 & 0.0020 & 0.0024 & 0.0025 \\ 	
		
		\bottomrule
	\end{tabular}
	\label{skew3}
\end{table}

\section{Appendix B}

\begin{table}[H]%[htb!]
	\centering
	\caption{Confidence interval index for $n=15$}
	\begin{adjustbox}{width=\textwidth}		
		\subfloat[$CV=0.1$]{
			\begin{tabular}{llll}
				\toprule
				& CP & CIL & $I$  \\ 
				\hline
				BSMill & 0.9090 & 0.0736 & 0.9658 \\ 
				MedMill & 0.9325 & 0.0759 & 0.9783 \\ 
				Mill & 0.9235 & 0.0746 & 0.9736 \\ 
				MedMMcK & 0.9445 & 0.0869 & 0.9831 \\ 
				MedMcK & 0.9445 & 0.0871 & 0.9830 \\ 
				MMcK & 0.9535 & 0.0854 & 0.9856 \\ 
				McK & 0.9535 & 0.0857 & 0.9856 \\ 
				S.K & 0.2120 & 0.0104 & 0.3789 \\ 
				C.P & 0.9155 & 0.0721 & 0.9696 \\ 
				BS C.P & 0.9025 & 0.0711 & 0.9626 \\ 
				Med C.P & 0.9265 & 0.0733 & 0.9755 \\ 
				Prop & 0.9520 & 0.0842 & 0.9861 \\ 
				Panich & 0.9510 & 0.0824 & 0.9865 \\ 
				NP.BS & 0.8395 & 0.0645 & 0.9273 \\ 
				PBS & 0.8700 & 0.0645 & 0.9452 \\
				\bottomrule
			\end{tabular}
		}\hfill
		\subfloat[$CV=0.3$]{
			\begin{tabular}{lll}
				\toprule
				CP & CIL & $I$ \\ 
				\hline
				0.9075 & 0.2411 & 0.9399 \\ 
				0.9295 & 0.2491 & 0.9514 \\ 
				0.9165 & 0.2437 & 0.9447 \\ 
				0.9435 & 0.2989 & 0.9528 \\ 
				0.9390 & 0.3072 & 0.9492 \\ 
				0.9505 & 0.2917 & 0.9573 \\ 
				0.9465 & 0.3015 & 0.9542 \\ 
				0.5235 & 0.0978 & 0.6953 \\ 
				0.9120 & 0.2354 & 0.9433 \\ 
				0.9015 & 0.2329 & 0.9376 \\ 
				0.9245 & 0.2406 & 0.9497 \\ 
				0.9280 & 0.2534 & 0.9500 \\ 
				0.9525 & 0.2791 & 0.9585 \\ 
				0.8400 & 0.2122 & 0.9040 \\ 
				0.8630 & 0.2125 & 0.9178 \\ 
				\bottomrule
			\end{tabular}
		}\hfill
		\subfloat[$CV=0.5$]{
			\begin{tabular}{llll}
				\toprule
				CP & CIL & $I$\\ 
				\hline
				0.9115 & 0.4631 & 0.9146 \\ 
				0.9295 & 0.4788 & 0.9235 \\ 
				0.9180 & 0.4678 & 0.9179 \\ 
				0.9540 & 0.7156 & 0.9110 \\ 
				0.9490 & 0.8401 & 0.9008 \\ 
				0.9545 & 0.6890 & 0.9133 \\ 
				0.9520 & 0.8265 & 0.9019 \\ 
				0.7265 & 0.2948 & 0.8178 \\ 
				0.9115 & 0.4519 & 0.9158 \\ 
				0.9060 & 0.4474 & 0.9131 \\ 
				0.9220 & 0.4626 & 0.9209 \\ 
				0.8915 & 0.4244 & 0.9071 \\ 
				0.9535 & 0.6405 & 0.9181 \\ 
				0.8650 & 0.4240 & 0.8910 \\ 
				0.8860 & 0.4239 & 0.9038 \\ 
				\bottomrule
			\end{tabular}
		}
	\end{adjustbox}	
	\label{A1}	
\end{table}

\begin{table}[H]%[htb!]
	\centering
	\caption{Confidence interval index for $n=25$}
	\begin{adjustbox}{width=\textwidth}	
		\subfloat[$CV=0.1$]{
			\begin{tabular}{llll}
				\toprule
				& CP & CIL & $I$  \\ 
				\hline
				BSMill & 0.9250 & 0.0559 & 0.9774 \\ 
				MedMill & 0.9345 & 0.0571 & 0.9824 \\ 
				Mill & 0.9305 & 0.0565 & 0.9803 \\ 
				MedMMcK & 0.9450 & 0.0617 & 0.9873 \\ 
				MedMcK & 0.9450 & 0.0618 & 0.9873 \\ 
				MMcK & 0.9500 & 0.0610 & 0.9901 \\ 
				McK & 0.9495 & 0.0611 & 0.9898 \\ 
				S.K & 0.2785 & 0.0101 & 0.4638 \\ 
				C.P & 0.9640 & 0.0714 & 0.9859 \\ 
				BS C.P & 0.9565 & 0.0707 & 0.9874 \\ 
				Med C.P & 0.9675 & 0.0722 & 0.9852 \\ 
				Prop & 0.9490 & 0.0603 & 0.9896 \\ 
				Panich & 0.9520 & 0.0598 & 0.9899 \\ 
				NP.BS & 0.8770 & 0.0513 & 0.9515 \\ 
				PBS & 0.8950 & 0.0512 & 0.9617 \\ 
				\bottomrule
			\end{tabular}
		}\hfill
		
		\subfloat[$CV=0.3$]{
			\begin{tabular}{llll}
				\toprule
				CP & CIL & $I$ \\  
				\hline
				0.9220 & 0.1827 & 0.9564 \\ 
				0.9340 & 0.1867 & 0.9625 \\ 
				0.9290 & 0.1843 & 0.9601 \\ 
				0.9445 & 0.2068 & 0.9656 \\ 
				0.9435 & 0.2094 & 0.9646 \\ 
				0.9440 & 0.2039 & 0.9657 \\ 
				0.9450 & 0.2070 & 0.9658 \\ 
				0.6775 & 0.0948 & 0.8163 \\ 
				0.9605 & 0.2331 & 0.9629 \\ 
				0.9620 & 0.2312 & 0.9628 \\ 
				0.9645 & 0.2362 & 0.9616 \\ 
				0.9235 & 0.1822 & 0.9573 \\ 
				0.9450 & 0.1989 & 0.9669 \\ 
				0.8745 & 0.1676 & 0.9313 \\ 
				0.8900 & 0.1675 & 0.9404 \\ 
				\bottomrule 
			\end{tabular}
		}\hfill
		
		\subfloat[$CV=0.5$]
		{
			\begin{tabular}{llll}
				\toprule
				CP & CIL & $I$  \\ 
				\hline
				0.9145 & 0.3476 & 0.9300 \\ 
				0.9300 & 0.3557 & 0.9380 \\ 
				0.9180 & 0.3503 & 0.9317 \\ 
				0.9430 & 0.4304 & 0.9367 \\ 
				0.9360 & 0.4505 & 0.9304 \\ 
				0.9515 & 0.4220 & 0.9414 \\ 
				0.9460 & 0.4454 & 0.9368 \\ 
				0.8240 & 0.2812 & 0.8840 \\ 
				0.9530 & 0.4431 & 0.9386 \\ 
				0.9560 & 0.4397 & 0.9383 \\ 
				0.9585 & 0.4500 & 0.9366 \\ 
				0.8795 & 0.3043 & 0.9149 \\ 
				0.9500 & 0.4070 & 0.9434 \\ 
				0.8825 & 0.3234 & 0.9142 \\ 
				0.8920 & 0.3232 & 0.9199 \\ 
				\bottomrule
		\end{tabular}}
	\end{adjustbox}	
	\label{A2}	
\end{table}

\begin{table}[H]%[htb!]
	\centering
	\caption{Confidence interval index for $n=50$}
	\begin{adjustbox}{width=\textwidth}			
		\subfloat[$CV=0.1$]{
			\begin{tabular}{lllll}
				\toprule
				& CP & CIL & $I$ \\ 
				\hline
				BSMill & 0.9365 & 0.0395 & 0.9863 \\ 
				MedMill & 0.9395 & 0.0400 & 0.9878 \\ 
				Mill & 0.9365 & 0.0398 & 0.9863 \\ 
				MedMMcK & 0.9475 & 0.0416 & 0.9918 \\ 
				MedMcK & 0.9475 & 0.0416 & 0.9918 \\ 
				MMcK & 0.9475 & 0.0413 & 0.9919 \\ 
				McK & 0.9475 & 0.0414 & 0.9919 \\ 
				S.K & 0.3740 & 0.0102 & 0.5714 \\ 
				C.P & 0.9950 & 0.0719 & 0.9804 \\ 
				BS C.P & 0.9945 & 0.0715 & 0.9806 \\ 
				Med C.P & 0.9950 & 0.0723 & 0.9804 \\ 
				Prop & 0.9455 & 0.0409 & 0.9909 \\ 
				Panich & 0.9465 & 0.0409 & 0.9914 \\ 
				NP.BS & 0.9080 & 0.0377 & 0.9711 \\ 
				PBS & 0.9110 & 0.0377 & 0.9728 \\ 
				\bottomrule
			\end{tabular}
		}\hfill
		\subfloat[$CV=0.3$]{
			\begin{tabular}{lll}
				\toprule
				CP & CIL & $I$ \\ 
				\hline
				0.9365 & 0.1273 & 0.9726 \\ 
				0.9445 & 0.1289 & 0.9767 \\ 
				0.9400 & 0.1281 & 0.9744 \\ 
				0.9515 & 0.1355 & 0.9784 \\ 
				0.9520 & 0.1362 & 0.9782 \\ 
				0.9520 & 0.1347 & 0.9785 \\ 
				0.9525 & 0.1355 & 0.9782 \\ 
				0.8290 & 0.0929 & 0.9162 \\ 
				0.9940 & 0.2316 & 0.9564 \\ 
				0.9930 & 0.2301 & 0.9568 \\ 
				0.9955 & 0.2330 & 0.9559 \\ 
				0.9320 & 0.1222 & 0.9709 \\ 
				0.9495 & 0.1331 & 0.9788 \\ 
				0.9090 & 0.1220 & 0.9581 \\ 
				0.9145 & 0.1220 & 0.9612 \\ 
				\bottomrule
			\end{tabular}
		}\hfill
		
		\subfloat[$CV=0.5$]{
			\begin{tabular}{llll}
				\toprule
				CP & CIL & $I$  \\ 
				\hline
				0.9380 & 0.2450 & 0.9567 \\ 
				0.9385 & 0.2478 & 0.9566 \\ 
				0.9340 & 0.2459 & 0.9544 \\ 
				0.9450 & 0.2726 & 0.9570 \\ 
				0.9430 & 0.2771 & 0.9553 \\ 
				0.9450 & 0.2703 & 0.9573 \\ 
				0.9455 & 0.2755 & 0.9569 \\ 
				0.9535 & 0.2794 & 0.9582 \\ 
				0.9925 & 0.4445 & 0.9293 \\ 
				0.9920 & 0.4428 & 0.9296 \\ 
				0.9935 & 0.4478 & 0.9287 \\ 
				0.8855 & 0.2065 & 0.9320 \\ 
				0.9495 & 0.2662 & 0.9604 \\ 
				0.9080 & 0.2365 & 0.9408 \\ 
				0.9210 & 0.2368 & 0.9482 \\
				\bottomrule
			\end{tabular}
		}
	\end{adjustbox}	
	\label{A3}	
\end{table}

\begin{table}[H]%[htb!]
	\centering
	\caption{Confidence interval index for $n=100$}
	\begin{adjustbox}{width=\textwidth}	
		\subfloat[$CV=0.1$]{
			\begin{tabular}{llll}
				\toprule
				& CP & CIL & $I$  \\ 
				\hline
				BSMill & 0.9530 & 0.0280 & 0.9948 \\ 
				MedMill & 0.9560 & 0.0282 & 0.9943 \\ 
				Mill & 0.9540 & 0.0281 & 0.9947 \\ 
				MedMMcK & 0.9630 & 0.0287 & 0.9930 \\ 
				MedMcK & 0.9630 & 0.0287 & 0.9930 \\ 
				MMcK & 0.9610 & 0.0286 & 0.9934 \\ 
				McK & 0.9610 & 0.0286 & 0.9934 \\ 
				S.K & 0.5355 & 0.0102 & 0.7232 \\ 
				C.P & 0.9995 & 0.0722 & 0.9796 \\ 
				BS C.P & 1.0000 & 0.0719 & 0.9796 \\ 
				Med C.P & 0.9995 & 0.0724 & 0.9796 \\ 
				Prop & 0.9600 & 0.0283 & 0.9936 \\ 
				Panich & 0.9615 & 0.0285 & 0.9933 \\ 
				NP.BS & 0.9380 & 0.0271 & 0.9892 \\ 
				PBS & 0.9405 & 0.0271 & 0.9905 \\ 
				\bottomrule
			\end{tabular}
		}\hfill
		
		\subfloat[$CV=0.3$]{
			\begin{tabular}{lll}
				\toprule
				CP & CIL & $I$  \\ 
				\hline
				0.9460 & 0.0907 & 0.9833 \\ 
				0.9520 & 0.0912 & 0.9850 \\ 
				0.9495 & 0.0910 & 0.9851 \\ 
				0.9480 & 0.0937 & 0.9839 \\ 
				0.9470 & 0.0939 & 0.9833 \\ 
				0.9475 & 0.0934 & 0.9837 \\ 
				0.9475 & 0.0937 & 0.9836 \\ 
				0.9510 & 0.0940 & 0.9848 \\ 
				1.0000 & 0.2337 & 0.9549 \\ 
				1.0000 & 0.2330 & 0.9550 \\ 
				1.0000 & 0.2344 & 0.9548 \\ 
				0.9280 & 0.0851 & 0.9744 \\ 
				0.9485 & 0.0928 & 0.9843 \\ 
				0.9350 & 0.0881 & 0.9778 \\ 
				0.9345 & 0.0883 & 0.9775 \\ 
				\bottomrule
			\end{tabular}
		}\hfill
		
		\subfloat[$CV=0.5$]{
			\begin{tabular}{llll}
				\toprule
				CP & CIL & $I$  \\ 
				\hline
				0.9385 & 0.1699 & 0.9674 \\ 
				0.9485 & 0.1715 & 0.9727 \\ 
				0.9430 & 0.1709 & 0.9698 \\ 
				0.9515 & 0.1813 & 0.9718 \\ 
				0.9520 & 0.1826 & 0.9716 \\ 
				0.9505 & 0.1805 & 0.9721 \\ 
				0.9510 & 0.1820 & 0.9718 \\ 
				0.9920 & 0.2724 & 0.9511 \\ 
				1.0000 & 0.4390 & 0.9283 \\ 
				1.0000 & 0.4365 & 0.9286 \\ 
				1.0000 & 0.4406 & 0.9281 \\ 
				0.8955 & 0.1416 & 0.9475 \\ 
				0.9525 & 0.1792 & 0.9719 \\ 
				0.9275 & 0.1667 & 0.9618 \\ 
				0.9315 & 0.1664 & 0.9641 \\ 
				\bottomrule
			\end{tabular}
		}
	\end{adjustbox}		
	\label{A4}		
\end{table}

\end{document}